\documentclass[12pt]{article}
\usepackage{amsfonts,amssymb,amsmath,amscd}
\usepackage[all,cmtip]{xy}
\usepackage{amsmath,amssymb,epsfig,ulem}

\newcommand{\ra}{\rightarrow}

\newcommand{\ZZ}{{\mathbb Z}}
\newcommand{\QQ}{{\mathbb Q}}
\newcommand{\RR}{{\mathbb R}}
\newcommand{\CC}{{\mathbb C}}
\newcommand{\TT}{{\mathbb T}}
\newcommand{\KK}{{\mathbb K}}

\renewcommand{\d}{{\rm d}}

\newcommand{\cH}{{\mathcal H}}

\renewcommand{\Im}{{\rm Im}}
\newcommand{\bz}{{\bar z}}

\newcommand{\cA}{{\mathcal A}}

\newcommand{\Hom}{{\rm Hom}}

\newcommand{\te}{{\tilde e}}
\newcommand{\D}{{\sf D}}
\newcommand{\sq}{{\mathsf q}}
\newcommand{\sM}{{\mathsf L}}

\newcommand{\Vect}{{\mathsf{Vect}}}

\newcommand{\Z}{\mathcal Z}

\newcommand{\End}{{\rm End}}

\newcommand{\tv}{{\tilde v}}

\newcommand{\tB}{{\tilde B}}

\newcommand{\hY}{{\hat Y}}

\newcommand{\g}{{\mathfrak t}}
\newcommand{\para}{{\parallel}}

\newcommand{\Mat}{{\rm Mat}}
\newcommand{\sA}{{\mathsf A}}
\newcommand{\sB}{{\mathsf B}}
\newcommand{\M}{{\Phi}}
\newcommand{\B}{{\varphi}}

\renewcommand{\sb}{{\mathsf g}}
\newcommand{\Y}{Q}
\newcommand{\hX}{{\hat X}}
\newcommand{\sbh}{{\hat{\mathsf g}}}

\def\be{\begin{equation}}
\def\ee{\end{equation}}
\def\bear{\begin{eqnarray}}
\def\eear{\end{eqnarray}}

\def\half{{ \frac{1}{2} }}

\def\Im{{\rm Im\hskip0.1em}}
\def\Id{{\mathbb{I}}}

\newcommand{\plusd}{\odot}

\def\a{{\alpha}}
\def\b{\beta}

\def\d{{\delta}}

\def\cC{{{\mathcal C}}}
\def\cL{{{\mathcal L}}}

\def\c{{\gamma}}

\def\p{{\partial}}

\def\vert{{|}}

\def\G{{\mathsf G}}

\title{Topological boundary conditions in abelian Chern-Simons theory}

\author{Anton Kapustin\\{\small \it California Institute of Technology, Pasadena, CA 91125,
U.S.A.}\\
Natalia Saulina\\{\small\it Perimeter Institute, Waterloo, Canada}}

\begin{document}

\begin{titlepage}

\maketitle

\begin{abstract}

We study topological boundary conditions in abelian Chern-Simons theory and line operators confined to such boundaries. From the mathematical point of view, their relationships are described by a certain 2-category associated to an even integer-valued symmetric bilinear form (the matrix of Chern-Simons couplings). We argue that boundary conditions correspond to Lagrangian subgroups in the finite abelian group classifying bulk line operators (the discriminant group). We describe properties of boundary line operators; in particular we compute the boundary associator. We also study codimension one defects (surface operators) in abelian Chern-Simons theories. As an application, we obtain a classification of such theories up to isomorphism, in general agreement with the work of Belov and Moore.
\end{abstract}

\vspace{-6in}
\parbox{\linewidth}
{\small\hfill \shortstack{pi-strings-195}} \vspace{6in}

\end{titlepage}

\section{Introduction}

Topological field theories provide an interesting playground for
studying the properties of path-integrals. Recently it became
clear that a complete understanding of TFT necessitates a study of
boundary conditions and defects of various dimensions. The
relationships between  topological defects are most conveniently
described using the machinery of higher category theory, see e.g. \cite{Kap:ICM} for an informal explanation of this. The goal of
this paper is to describe some of this machinery in the simplest
nontrivial example of a 3d TFT: abelian Chern-Simons theory. Any 3d
TFT associates a number  to a closed
oriented 3-manifold, a vector space to a Riemann
surface, a category to a circle, and a
2-category to a point. In the case of Chern-Simons theory the first
three of these are well understood; our goal (which is partially
achieved in this paper) is to describe the last one.

The case of abelian Chern-Simons theory is particularly attractive because its action is completely determined by a simple datum: an even integral symmetric bilinear form $K$. Thus from a mathematical point of view the path-integral for Chern-Simons theory can be viewed as a black box which produces a 2-category from such a form. Abelian Chern-Simons theories are also known to describe some phases of actual physical systems (Fractional Quantum Hall phases). We will see that not all abelian Chern-Simons theories admit topological boundary conditions; in particular, the set of boundary conditions appears to be nonempty only if $K$ has zero signature.\footnote{We stress that we only study topological boundary conditions. It is well-known that any Chern-Simons theory admits a boundary which carries a chiral WZW model; however, these degrees of freedom are not topological (the partition function of Chern-Simons theory coupled to such boundary degrees of freedom depends on the conformal class of the metric on the boundary).} For such $K$ basic boundary conditions on the classical level are labeled by Lagrangian subgroups of the gauge group $\TT$, where ``Lagrangian'' means isotropic and coisotropic with respect to an indefinite metric on $\TT$ given by the bilinear form $K$. On the quantum level the answer is similar, but the role of $\TT$ is played by a certain finite abelian group $\D$ (the discriminant group), and the role of $K$ is played by a $\QQ/\ZZ$-valued quadratic form $\sq$ on $\D$.  Both of these are determined by $K$. Objects of the 2-category of boundary conditions are labeled by Lagrangian subgroups of the group $\D$.

Different symmetric bilinear forms may give rise to identical pairs $(\D,\sq)$. We will see that many other properties of the theory depend only on $(\D,\sq)$, and this suggests that many abelian Chern-Simons theories which are distinct on the classical level are isomorphic quantum-mechanically. We will argue below that isomorphism classes of quantum abelian Chern-Simons theories correspond to stable equivalence classes of even integer-valued symmetric bilinear forms. This is in general agreement with the work of Belov and Moore \cite{BM}; we discuss the precise relationship below. Our approach to establishing an isomorphism between TFTs is based on the notion of a ``duality wall'' \cite{GW,KapTi}. This is a surface operator which separates two isomorphic theories; it is distinguished among all such surface operators by the property of being invertible.

Since abelian Chern-Simons theory is a free theory, there are no perturbative quantum corrections, and one might naively think that quantum effects are relatively trivial. We will see that this expectation is wrong because of nonperturbative quantum effects due to so-called monopole operators. The effects of these operators are important even for bulk observables (Wilson lines); for example, we will see that they lead to a nontrivial associator in the Operator Product of Wilson lines. The mathematical counterpart of this fact is well-known and goes back to the work of Joyal and Street \cite{JS}, but as far as we know until now there was no physical derivation of these results.

Let us describe the content of the paper. In section
\ref{sec:review} we briefly review abelian Chern-Simons theory and
its observables. In section \ref{sec:bulklines} we discuss bulk line
operators and their Operator Product Expansion (OPE), which is
encoded in a braided monoidal structure on the category of bulk line
operators. We will see that a crucial role is played by monopole operators. In section \ref{sec:boundaries} we discuss simple
boundary conditions for abelian Chern-Simons theory which do not
involve boundary degrees of freedom. In section \ref{sec:bdrylines}
we study properties of boundary line operators and
boundary-changing line operators. In particular, we determine the
monoidal structure on the category of boundary line operators. In
section \ref{sec:generalbdries} we study more general boundary
conditions which involve topological boundary degrees of freedom.
 In
section \ref{sec:surface} we discuss surface operators in abelian
Chern-Simons theories and apply the results to the classification of
abelian Chern-Simons theories on the quantum level. As mentioned
above, we find that two Chern-Simons theories are isomorphic if the corresponding bilinear forms have the same signature and
are stably equivalent. In section \ref{sec:concl} we summarize our
results and formulate some natural conjectures regarding the
2-category of boundary conditions in abelian Chern-Simons theory. In the appendices we review some basic notions pertaining to monoidal categories. This material is standard and is included here because it is unfamiliar to most physicists.

\section{Review of abelian Chern-Simons theory}\label{sec:review}

\subsection{Action and observables}

The gauge group of an abelian Chern-Simons theory is a torus which we will think of as a quotient of an $n$-dimensional vector space by a subgroup $\Lambda\simeq \ZZ^n$. We will denote this torus $\TT_\Lambda$. In general, for any finite-rank free abelian group $\Gamma$ we will denote by $\TT_\Gamma$ the torus $(\Gamma\otimes\RR)/(2\pi\Gamma)$. The gauge field is a connection on a principal $\TT_\Lambda$-bundle over an oriented 3-manifold $M_3$; locally, on a trivializing patch $U\subset M_3$, it is represented by a 1-form $A=A_{\mu} dx^{\mu}$ taking values in the vector space $t_{\Lambda}=\Lambda\otimes\RR,$ the Lie algebra of $\TT_\Lambda.$
Under a gauge transformation $\alpha:U\ra \TT_\Lambda$ it transforms as
$$
A\mapsto A+d \alpha.
$$
The action in Euclidean signature is schematically
$$
S_{CS}=\frac{i}{4\pi}\int_{M_3} K(A,dA),
$$
where $K$ is a symmetric bilinear form on $\Lambda\otimes\RR$. It depends only on the smooth structure on $M_3$. In order for $\exp(-S)$ to be well-defined (i.e. independent of the choice of trivialization of the $\TT_\Lambda$-bundle), $K$ must be integer-valued and even on $\Lambda$. That is, for any $\lambda,\lambda'\in\Lambda$ we must have
$$
K(\lambda,\lambda')\in\ZZ,\quad K(\lambda,\lambda)\in 2\ZZ.
$$
A more careful analysis shows that one can make sense of the Chern-Simons theory even if $K$ is odd, provided we endow $M_3$ with a spin structure. Such a variant of the theory is called spin Chern-Simons theory; it has been studied in detail by Belov and Moore \cite{BM}. We will not consider spin Chern-Simons theories in this paper.

Following the mathematical terminology, we will call a free abelian group $\Lambda\simeq \ZZ^n$ equipped with an integer-valued symmetric bilinear form $K$ a lattice of rank $n$. We will say that a lattice is even if $K$ is even. Thus classical abelian Chern-Simons theories are labeled by even lattices.

The basic observables in abelian Chern-Simons theory are Wilson loops. Let $X$ be an element of the dual lattice $\Lambda^*=\Hom(\Lambda,\ZZ)$, i.e. a linear function on $\Lambda\otimes\RR$ which takes integer values on $\Lambda$. Let $L$ be an oriented closed curve in $M_3$. The Wilson loop with charge $X$ and support $L$ is
$$
W_X(L)=\exp\left(-i\int_L X(A)\right).
$$
In other words, $X$ defines a unitary one-dimensional representation of $\TT_\Lambda$, $X(A)$ is the connection on the associated line bundle, and $W_X(L)$ is the holonomy of this connection along $L$.

Chern-Simons theory also admits observables localized on surfaces (surface operators); we will discuss them
in section \ref{sec:surface}.

\subsection{Space of states}

Any 3d TFT associates an inner-product vector space $\cH_\Sigma$ to a Riemann surface $\Sigma$. The mapping class group of $\Sigma$ (i.e. the quotient of the group of diffeomorphisms of $\Sigma$ by its identity component) acts projectively on $\cH_\Sigma$. In the case of abelian Chern-Simons theory $\cH_\Sigma$ is obtained by geometric quantization of the moduli space of flat $\TT_\Lambda$-connections on $\Sigma$. The latter space is a torus with a symplectic form
$$
\omega=\frac{1}{4\pi}\int_\Sigma K(\delta A,\delta A).
$$
Its quantization is the space of holomorphic sections of a line
bundle $\cL$ whose curvature is $\omega$. For a genus $g$ Riemann
surface $\Sigma,$ it has dimension $\vert \det K\vert^g$. As discussed in \cite{BM}, the action of
the mapping class group of $\Sigma_g$ on $\cH_\Sigma$ factors through the group $Sp(2g,\ZZ)$. Explicit formulas for the action of $Sp(2g,\ZZ)$ on $\cH_\Sigma$ are given in \cite{BM}; we will not need them in this paper.

\section{Line operators in the bulk}\label{sec:bulklines}

\subsection{Monopole identifications}

In any 3d TFT observables localized on closed curves form an additive $\CC$-linear category which we will call the category of line operators. Given a pair of line operators, the space of morphisms between them is the space of local operators inserted at their joining point. Equivalently, the space of morphisms is the vector space which 3d TFT attaches to a 2-sphere with two marked points, the points being marked by line operators.

In the case of abelian Chern-Simons theory line operators are Wilson lines $W_X$, $X\in\Lambda^*$, and their direct sums. A Wilson line is localized on a one-dimensional submanifold of $M_3$ which may have boundary points. The only local operator which can be inserted into a given Wilson loop $W_X$ is the identity operator (times an arbitrary complex number). Thus the space of endomorphisms of $W_X$ is $\CC$, for any $X$.\footnote{The category of bulk line operators is in fact semi-simple, and Wilson lines are simple objects in this category.}

Local operators sitting at the joining point of two distinct Wilson lines $W_X$ and $W_{X'}$ are more interesting. Such an operator  must carry electric charge $X'-X$. To construct such an operator we need to consider an insertion of a Dirac monopole. Recall that a Dirac monopole is characterized by the topological class of the gauge field on a small 2-sphere centered on it. The topology of a $\TT_\Lambda$ bundle on $S^2$ is described by its first Chern class
$$
m=\frac{1}{2\pi}\int_{S^2} dA \in \Lambda.
$$
In Chern-Simons theory such an operator is not gauge-invariant: under a gauge transformation $\alpha:M_3\ra\TT_\Lambda$ the exponential of minus the Chern-Simons action is multiplied by a phase
$$
\exp(-i K(\alpha(p),m)),
$$
where $p$ is the insertion point of the monopole operator. This means that the electric charge of a Dirac monopole is $Km$ with $m\in\Lambda$. Such an operator can be inserted at the joining point of $W_X$ and $W_{X'}$ provided $X'-X=Km$. Thus $\Hom(W_X,W_{X'})=\CC$ if $X'-X=Km$ for some $m\in\Lambda$, and zero otherwise.

A nonzero morphism between $W_X$ and $W_{X+Km}$ is invertible (its inverse is the Dirac monopole with magnetic charge $-m$). Thus line operators $W_X$ and $W_{X+Km}$ are isomorphic as objects of the category of line operators. It is convenient to identify isomorphic objects, thereby passing to an equivalent category whose objects are isomorphism classes of Wilson line observables. The nice thing about abelian Chern-Simons theory is that after such identification the set of objects is finite. Namely, it is a finite abelian group $\D=\Lambda^*/\Im\, K$, where $\Im\, K$ is the image of $\Lambda$ in $\Lambda^*$ under the map $K:\Lambda\ra \Lambda^*$. The group $\D$ is called the discriminant group of the lattice $(\Lambda,K)$. From now on we will regard the charge $X$ as an element of the quotient group $\D=\Lambda^*/\Im\, K$.

If the inverse of $K$ is integral, then $\Im\, K=\Lambda^*$ and the group $\D$ is trivial. Such a lattice $(\Lambda,K)$ is called self-dual in the physical literature and unimodular in the mathematical literature.  It can be shown that the signature of an even unimodular lattice is divisible by $8$. An example of a positive-definite even unimodular lattice is the root lattice of $E_8$; in view of the above the corresponding Chern-Simons theory has trivial discriminant group and therefore all Wilson line observables are isomorphic to the trivial one. Another example of an even unimodular lattice is $\ZZ^2$ with an indefinite form
$$
K=\begin{pmatrix} 0 & 1 \\ 1 & 0\end{pmatrix}.
$$
The corresponding lattice $(\Lambda,K)$ is often denoted $U$ in the mathematical literature.
The corresponding Chern-Simons theory has gauge group $U(1)\times U(1)$ and an action
$$
\frac{i}{4\pi}\int_M (adb+bda).
$$
In this theory all Wilson lines are trivial as well.

To get an example of a Chern-Simons theory with a nontrivial category of line operators we can take $G=U(1)$ and let $K=k\in 2\ZZ$. Then $\D=\ZZ_k$.

\subsection{Braided monoidal structure}

The category of line operators in any 3d TFT is a semi-simle braided monoidal category.\footnote{In fact, a ribbon category, see e.g. \cite{JS,BK,Sti} for a definition.} Very roughly, a monoidal category is a category with an associative tensor product; a braided monoidal category is a monoidal category in which there is a specified isomorphism between $A\otimes B$ and $B\otimes A$ for any two objects $A$ and $B$. Precise definitions can be found in appendix A.

For abelian Chern-Simons theory the braided monoidal structure can be described very explicitly \cite{Sti}. The tensor product of objects arises from the Operator Product of Wilson lines, which implies that
$$
W_X\otimes W_{Y}= W_{X+Y},\quad X,Y\in\D.
$$
The definition of monoidal structure also involves an associator which is an isomorphism
$$
a(X,Y,Z): (W_X\otimes W_Y)\otimes W_Z\ra
W_X\otimes (W_Y\otimes W_Z).
$$
This isomorphism must satisfy the pentagon identity, see appendix A. In the case of abelian Chern-Simons theory the associator is merely a nonzero complex number depending on $X,Y$ and $Z$, but it is nevertheless fairly nontrivial. To see how this nontrivial associator comes about, let us first look at the braiding. The braiding is an isomorphism
$$
W_X\otimes W_Y\ra W_Y\otimes W_X,\quad X,Y\in\D,
$$
satisfying a compatibility condition with the associator (the hexagon identities, see appendix A). In the case of abelian Chern-Simons theory the braiding isomorphism originates from the Aharonov-Bohm phase which arises when two Wilson lines arranged on a line exchange order. This phase is
$$
\exp(-i\pi K^{-1}(X,Y)).
$$
Here $X,Y\in\Lambda^*$ are the electric charges of the two Wilson lines, and $K^{-1}$ is the $\QQ$-valued bilinear form on $\Lambda^*$ whose matrix is the inverse of that of $K$.

It will be convenient to use additive notation for Aharonov-Bohm phases, i.e. we will think of them as elements of $\QQ/\ZZ$ rather than of $U(1)$ and will write $\phi$ instead of $\exp(2\pi i\phi)$. Thus we will write the braiding phase as
\begin{equation}\label{bulkbraiding}
s(X,Y)=-\frac12 K^{-1}(X,Y).
\end{equation}
The key observation is that this phase is not invariant under $X\mapsto X+Km$ but may shift by $1/2$ (in the multiplicative notation this corresponds to a sign ambiguity). Thus the braiding isomorphism is not a well-defined function on $\D\times \D$. To fix the ambiguity, one has to choose particular representatives in $\Lambda^*$ for all elements in $\D=\Lambda^*/\Im\, K$.

It is this innocent-looking procedure that generates a nontrivial
associator. The simplest way to see this is to look at the hexagon
identities (appendix A) relating the associator and the braiding. The braiding
phase $s(X,Y)$ is a $\QQ/\ZZ$-valued function on
$\D\times \D$, and the hexagon identities (\ref{hexbulk1},\ref{hexbulk2}) say that the failure of $s$
to be bilinear is expressed through the associator. The braiding
phase (\ref{bulkbraiding}) is bilinear when regarded as a function
on $\Lambda^*\times\Lambda^*$, so if we do not identify Wilson lines
whose charges differ by an element of $\Im\, K$, it is consistent to
set the associator to zero. But to regard it as a function on
$\D\times\D$, we need to choose particular representatives in
$\Lambda^*$ for all elements of $\D$, and then the braiding
(\ref{bulkbraiding}) is no longer bilinear, and the associator must
be nontrivial.

Let us now compute the associator for bulk Wilson lines in
Chern-Simons theory. Consider first a simple example where
$\D=\ZZ_M$ for some $M$. Let $e\in\Lambda^*$ be a generator of $\D$,
then we can uniquely lift an element $X\in\D$ to an element of
$\Lambda^*$ of the form $A e$, where $A$ is an integer from $0$ to
$M-1$. That is, we can label Wilson lines by integers from $0$ to
$M-1$. Imagine three Wilson lines $W_A,W_B,W_C$ with charges $A,B,C\in [0,M-1]$
extended in the time direction and placed along the $x$-axis in
$\RR^3$ in this order. Whenever the merger of two Wilson lines
produces a Wilson line with charge $U e\in\Lambda^*$ with $U\geq M$, we must
convert $U$ to the range $[0,M-1]$ by attaching a semi-infinite
Wilson line terminating on a monopole operator with a magnetic
charge $K^{-1}Me$. We will move these semi-infinite Wilson lines to
the left of the left-most Wilson line $W_X$. Then it is easy to see
that the difference between two different ways of computing the
triple product $W_A\otimes W_B\otimes W_C$ is the braiding of $W_A$
and the semi-infinite Wilson line arising from the merger of $W_B$
and $W_C$. Thus the associator $a(A,B,C)=\exp(2\pi i h(A,B,C))$ is a phase given by
$$
h(A,B,C)=\frac12 A \left[{B+C\over M}\right] M K^{-1}(e,e),
$$
where $[w]$ stands for integral part of $w.$
Note that the bilinear form $MK^{-1}$ on $\Lambda^*$ is integral, so from the multiplicative viewpoint the associator is at most a sign.\footnote{It is easy to show that for odd $M$ the form $M K^{-1}$ is even, so the associator can be nontrivial only for even $M$.} A straightforward computation best left to the reader shows that $h(A,B,C)$ satisfies the pentagon identity. The braiding in this case is
$$
s(A,B)=-\frac12 AB\, K^{-1}(e,e).
$$
It is easy to verify that the hexagon identities are also satisfied.

The general case is surprisingly subtle. Suppose the discriminant group is
$$
\D\simeq \bigoplus_{i=1}^N \ZZ_{M_i}.
$$
Let $e_1,\ldots,e_N\in\Lambda^*$ be generators of $\D$. This means
that any element of $\Lambda^*$ can be written in a unique way as a
linear combination of $e_i$ with integral coefficients plus an
element in $\Im K$, and in addition $M_i e_i\in\Im K$ for all $i$.
Then we can uniquely lift $X\in\D$ to an element of $\Lambda^*$ of
the form
$$
A_1 e_1+\cdots +A_N e_N,
$$
where $A_i$ is an integer in the  range $[0,M_i-1]$. A Wilson line
is labeled by an integer vector $\vec{A}=(A_1,\ldots,A_N)$ with all
components in the standard range.  The group operation on such
vectors will be denoted $\vec{A}\plusd \vec{B}$ and is given
explicitly by
$$
(\vec{A}\plusd \vec{B})_i=(A_i+B_i)\,\, {\rm mod} \,\, M_i.
$$

To compute the braiding and the associator, we first need to define more precisely Wilson lines and their tensor product. Let us choose an order on the set of generators of $\D$:
$$e_1 < e_2 < \ldots < e_{N-1} < e_N.$$
We define a Wilson line $W_{\vec{A}}$ as the ordered product of
Wilson lines $W_{A_1 e_1},\ldots, W_{A_N e_N}.$  This makes sense if
we think of a Wilson line as a thin ribbon oriented in a particular
direction. For brevity we will refer to $W_{A_ie_i}$ as constituents of
$W_{\vec{A}}$.

\vspace{1cm}

\setlength{\unitlength}{3cm}
\begin{picture}(1,1)
 \linethickness{0.3mm}
\put(2, 0.145){\line(0, 1){1}} \put(2, 0){$W_{\vec{A}}$} \put(2.3,
0.5){$=$}
 \multiput(2.7, 0.145)(0.1,0){7}{\line(0, 1){1}}
 \put(2.5, 0){$W_{A_1e_1}$}
 \put(3, 0){$\ldots $}
  \put(3.3, 0){$W_{A_Ne_N}$}
 \end{picture}

 \vspace{1cm}

Naively, the tensor product of two Wilson lines
$W_{\vec{A}},W_{\vec{B}}$ is defined by placing them side by side
along the chosen direction and fusing. The expected answer is
$W_{\vec{A}\plusd\vec{B}}$, but to see this we need to modify the
naive definition in two ways. First, we need to rearrange the
constituents to put them in the standard order. Second, in general
the $i^{\rm th}$ component of $\vec{A}+\vec{B}$ is  outside of the
standard range $[0,M_i-1]$. To bring it to the standard range we
write
$$
A_i+B_i=(A_i\plusd B_i)+M_i \left[\frac{A_i+B_i}{M_i}\right],
$$
and move the  Wilson lines
$I_{(A_i,B_i)}$ with charges
$$
e_i M_i \left[\frac{A_i+B_i}{M_i}\right]
$$
to the left of all other constituent Wilson lines. The net result is
the Wilson line $W_{\vec{A}\plusd\vec{B}}$ together with the Wilson
line $I_{(\vec A,\vec B)}$ with the charge
\begin{equation}\label{trivW}
\sum_i e_i M_i \left[\frac{A_i+B_i}{M_i}\right]
\end{equation}
to the left of it. Since (\ref{trivW}) lies in $\Im K$, the Wilson
line $I_{(\vec A,\vec B)}$ is isomorphic to the trivial one. The
figure below illustrates the tensor multiplication of $W_{\vec{A}}$
and $W_{\vec{B}}$ for $N=2.$

  \vspace{1cm}

\setlength{\unitlength}{3cm}
\begin{picture}(1,1)
 \linethickness{0.3mm}
\put(0, 0.145){\line(0, 1){1}} \put(0.2, 0.145){\line(0, 1){1}}
\put(-0.2, 0){$W_{A_1}$} \put(0.1, 0){$W_{A_2}$}

\put(0.5, 0.145){\line(0, 1){1}} \put(0.7, 0.145){\line(0, 1){1}}
\put(0.4, 0){$W_{B_1}$} \put(0.7, 0){$W_{B_2}$}

\put(0.8, 0.5){$\underrightarrow{e^{-i\pi A_2 \, B_1 \sb_{21}}}$}

\put(1.5, 0.145){\line(0, 1){1}} \put(1.8, 0.145){\line(0, 1){1}}
\put(1.3, 0){$I_{(A_1,B_1)}$} \put(1.8, 0){$W_{A_1\plusd B_1}$}

\put(2.5, 0.145){\line(0, 1){1}} \put(2.8, 0.145){\line(0, 1){1}}
\put(2.3, 0){$I_{(A_2,B_2)}$} \put(2.8, 0){$W_{A_2\plusd B_2}$}

\put(2.9, 0.5){$\underrightarrow{e^{-i\pi M_2 \, [{A_2+B_2\over
M_2}]\,(A_1\plusd B_1)\, \sb_{21}}}$}

\put(4.5, 0.145){\line(0, 1){1}} \put(5, 0.145){\line(0, 1){1}}
\put(4.4, 0){$I_{(\vec A, \vec B)}$}

% \put(5, 0.145){\line(0, 1){1}}\put(5.2, 0.145){\line(0, 1){1}}
 \put(5, 0){$W_{\vec A \plusd \vec B}$}

 \end{picture}

 \vspace{1cm}

 It should be noted that there is a certain arbitrariness in this definition. We are free to redefine the tensor product
 $$
 W_{\vec A}\otimes W_{\vec B}\ra W_{{\vec A}\plusd {\vec B}}
 $$
 by an arbitrary invertible endomorphism of  $W_{{\vec A}\plusd {\vec B}}$. Since the endomorphism algebra of any Wilson line is $\CC$, we may regard this endomorphism as a phase $k(\vec A,\vec B)$. This modifies both the braiding and the associator in a fairly obvious way, see appendix A for details. For example the braiding is modified as follows:
 $$
 s(\vec A,\vec B)\ra s(\vec A,\vec B)-k(\vec A,\vec B)+k(\vec B,\vec A).
 $$
 We will say that such a redefinition modifies the associator and the braiding by a coboundary. Braided monoidal categories whose associator and braiding differ by a coboundary are regarded as equivalent.

We are now ready to compute the braiding and the associator.
Naively, braiding is a particular isomorphism of $W_{\vec{A}}\otimes
W_{\vec{B}}$ and $W_{\vec{B}}\otimes W_{\vec{A}}$ arising from
moving the second Wilson counterclockwise around the first one. The
corresponding Aharonov-Bohm phase is
$$
s_0(\vec A,\vec B)=-\frac12 \sum_{i,j} A_i B_j \sb_{ij},
$$
where we denoted $\sb_{ij}=K^{-1}(e_i,e_j)$. However, since our
definition of the tensor product is more complicated than mere
fusion, one needs to correct this expression. The tensor product
operation involves rearranging the constituents of Wilson lines and
then splitting off Wilson lines isomorphic to the trivial one and
moving them to the left of all other constituents. The Aharonov-Bohm
phase associated with the first step of this operation is
$$
\psi_1(\vec{A},\vec{B})=-\frac12 \sum_{i>j} A_i B_j \sb_{ij}.
$$
The Aharonov-Bohm phase associated with the second step of the operation is
$$
\psi_2(\vec{A},\vec{B})=-\frac12 \sum_{i>j} M_i
\left[\frac{A_i+B_i}{M_i}\right](A_j\plusd B_j)\sb_{ij}.
$$
The phase from both steps is
$$
\psi(\vec{A},\vec{B})=\psi_1(\vec{A},\vec{B})+\psi_2(\vec{A},\vec{B}).
$$
The braiding is an invertible endomorphism of
$W_{\vec{A}\plusd\vec{B}}$ corresponding to the following
manipulations: first we reverse the above two steps to get from $
W_{\vec{A}\plusd\vec{B}}$ to a composite of $W_{\vec{A}}$ and
$W_{\vec{B}}$, then we exchange the order of $W_{\vec{A}}$ and
$W_{\vec{B}}$ by moving $W_{\vec{B}}$ counterclockwise, and finally
we reassemble $I_{(\vec A,\vec B)}\, W_{\vec{A}\plusd\vec{B}}$.

 \vspace{1cm}

\setlength{\unitlength}{3cm}
\hspace{2cm} \begin{picture}(1,1)
 \linethickness{0.3mm}
 \put(-0.2, 0.145){\line(0, 1){1}} \put(-0.3, 0){$I_{(\vec A, \vec B)}$}
\put(0.1, 0.145){\line(0, 1){1}} \put(0.1, 0){$W_{\vec A \plusd \vec
B}$} \put(0.2, 0.5){$\mapsto e^{-i\pi \psi(\vec A,\vec B)}$} \put(1,
0.145){\line(0, 1){1}} \put(1.3, 0.145){\line(0, 1){1}} \put(1,
0){$W_{\vec A}$} \put(1.3, 0){$W_{\vec B}$} \put(1.5, 0.5){$\mapsto
e^{i\pi s_0(\vec A,\vec B)}$} \put(2.5, 0.145){\line(0, 1){1}}
\put(2.8, 0.145){\line(0, 1){1}} \put(2.5, 0){$W_{\vec B}$}
\put(2.8, 0){$W_{\vec A}$} \put(3.1, 0.5){$\mapsto e^{i\pi \psi(\vec
B,\vec A)}$}  \put(4, 0.145){\line(0, 1){1}} \put(3.9, 0){$I_{(\vec
A, \vec B)}$}\put(4.3, 0.145){\line(0, 1){1}} \put(4.3, 0){$W_{\vec
A\plusd \vec B}$}
 \end{picture}

 \vspace{1cm}

The corresponding Aharonov-Bohm phase is
$$
s(\vec{A},\vec{B})=s_0(\vec{A},\vec{B})
-\psi(\vec{A},\vec{B})+\psi(\vec{B},\vec{A}).
$$
Note that the phase $\psi_2$ drops out of this expression because it
is symmetric in $\vec{A}$ and $\vec{B}$. Therefore the braiding
phase is
$$
s(\vec{A},\vec{B})=-\sum_{i<j} A_i B_j \sb_{ij}-\frac12 \sum_i A_i B_i
\sb_{ii}.
$$
This agrees with the expression in \cite{Sti} and differs from the
naive $s_0$ by a coboundary
$-\psi(\vec{A},\vec{B})+\psi(\vec{B},\vec{A})$. We will see shortly
that this choice of a coboundary simplifies the associator.

Now let us turn to computing the associator $a(A,B,C)=\exp(2\pi i h(A,B,C))$. This is a particular
isomorphism from $(W_{\vec{A}}\otimes W_{\vec{B}})\otimes
W_{\vec{C}}$ to $W_{\vec{A}}\otimes (W_{\vec{B}}\otimes
W_{\vec{C}})$. Both tensor products give
$W_{\vec{A}\plusd\vec{B}\plusd\vec{C}}$ together with some Wilson
lines isomorphic to the trivial one. We illustrate this below.

 \vspace{1cm}

\setlength{\unitlength}{3cm} \hspace{2cm} \begin{picture}(1,1)
 \linethickness{0.3mm}
\put(0.1, 0.145){\line(0, 1){1}} \put(0.1, 0){$W_{\vec A}$}
\put(0.3, 0.145){\line(0, 1){1}} \put(0.3, 0){$W_{\vec B}$}
\put(0.6, 0.145){\line(0, 1){1}} \put(0.6, 0){$W_{\vec C}$}
 \put(0.7,
0.5){$\underrightarrow{e^{2\pi i \psi(\vec A, \vec B)}}$}
 \put(1.6,0.145){\line(0, 1){1}} \put(1.9, 0.145){\line(0, 1){1}}
 \put(2.4, 0.145){\line(0, 1){1}}

\put(1.5, 0){$I_{(\vec A,\vec B)}$} \put(1.9, 0){$W_{\vec A \plusd
\vec B}$}\put(2.4, 0){$W_{\vec C}$}

\put(2.5, 0.5){$\underrightarrow{e^{2\pi i (\psi(\vec A\plusd \vec
B,\vec C)+h_1)}}$}

\put(3.5, 0.145){\line(0, 1){1}} \put(3.5, 0){$I_{(\vec A,\vec
B,\vec C)}$} \put(3.9,0.145){\line(0, 1){1}} \put(4, 0){$W_{\vec A
\plusd \vec B \plusd \vec C}$}
\end{picture}

 \vspace{2cm}

\setlength{\unitlength}{3cm} \hspace{2cm} \begin{picture}(1,1)
 \linethickness{0.3mm}
\put(0.1, 0.145){\line(0, 1){1}} \put(0.1, 0){$W_{\vec A}$}
\put(0.5, 0.145){\line(0, 1){1}} \put(0.5, 0){$W_{\vec B}$}
\put(0.7, 0.145){\line(0, 1){1}} \put(0.7, 0){$W_{\vec C}$}
 \put(0.8,
0.5){$\underrightarrow{e^{2\pi i \psi(\vec B, \vec C)}}$}
 \put(1.6,0.145){\line(0, 1){1}} \put(1.9, 0.145){\line(0, 1){1}}
 \put(2.4, 0.145){\line(0, 1){1}}

\put(1.5, 0){$W_{\vec A}$} \put(1.9, 0){$I_{(\vec B, \vec
C)}$}\put(2.4, 0){$W_{\vec B\plusd \vec C}$}

\put(2.5, 0.5){$\underrightarrow{e^{2\pi i (\psi(\vec A, \vec
B\plusd \vec C)+h_2)}}$}

\put(3.5, 0.145){\line(0, 1){1}} \put(3.5, 0){$I_{(\vec A,\vec
B,\vec C)}$} \put(3.9,0.145){\line(0, 1){1}} \put(4, 0){$W_{\vec A
\plusd \vec B \plusd \vec C}$}

 \end{picture}

 \vspace{1cm}

The extra phases $h_1$ and $h_2$ arise since``trivial'' Wilson lines
have a nontrivial braiding with other Wilson lines and between
themselves. To compare the two tensor products we need to bring them
to the standard position, i.e. to the left of all constituents of
$W_{\vec{a}\plusd\vec{b}\plusd\vec{c}}$ and arranged in the order of
increasing $i$. Concretely, the phase $h_1$  arises from the
rearrangement of ``trivial'' Wilson lines from $I_{(\vec A,\vec
B)}I_{(\vec A \plusd \vec B, \vec C)}$ to $I_{(\vec A,\vec B,\vec
C)}$ by moving $I_{(A_j \plusd B_j, C_j)}$ counterclockwise to the
left of $I_{(A_i,B_i)}$ for $i > j$:
$$
h_1(\vec{A},\vec{B},\vec{C})=-\frac12 \sum_{i>j} M_i
\left[\frac{A_i+B_i}{M_i}\right]M_j \left[\frac{(A_j\plusd
B_j)+C_j}{M_j}\right] \sb_{ij}.
$$

Meanwhile, the phase $h_2$ arises by first moving $W_{\vec A}$
clockwise to the right of $I_{(\vec B,\vec C)}$ and then
rearranging trivial lines from $I_{(\vec B,\vec C)}I_{(\vec A, \vec
B\plusd \vec C)}$ to $I_{(\vec A,\vec B,\vec C)}$ by moving
$I_{(A_j, B_j \plusd C_j)}$ counterclockwise to the left of
$I_{(B_i,C_i)}$ for $i > j$:
$$
h_2(\vec{A},\vec{B},\vec{C})=\frac12 \sum_{i,j} A_i
M_j\left[\frac{B_j+C_j}{M_j}\right]\sb_{ij}-\frac12\sum_{i>j} M_i
\left[\frac{B_i+C_i}{M_i}\right]M_j \left[\frac{(B_j\plusd
C_j)+A_j}{M_j}\right] \sb_{ij}.
$$
So the associator is given by
$$h(\vec A,\vec B,\vec C)=h_2(\vec A,\vec B,\vec C)-h_1(\vec A,\vec B,\vec C)
 +\psi(\vec{B},\vec{C})+
\psi(\vec{A},\vec{B}\plusd\vec{C})-\psi(\vec{A},\vec{B})-\psi(\vec{A}\plusd\vec{B},\vec{C}).$$
This differs from the ``naive'' associator $h_2-h_1$  by a
coboundary term.

After a somewhat lengthy computation we get a remarkably simple
associator
$$
h(\vec{A},\vec{B},\vec{C})=\frac12\sum_i M_i A_i
\left[\frac{B_i+C_i}{M_i}\right]\sb_{ii}
$$
To arrive at this result, we repeatedly used the fact that $M_i \sb_{ij}$ (no summation is implied) is integral. Indeed, since $M_i e_i=K f_i$ for some $f_i\in\Lambda$, we have
$$
M_i \sb_{ij}=K^{-1}(Kf_i,e_j)=(f_i,e_j)\in\ZZ.
$$
One can also check that $h$ satisfies the pentagon identity, and $h$
and $s$ together satisfy the hexagon identities.

The braided monoidal structure described above depends on far less data than the even symmetric bilinear form $K$. To describe what it really depends on, note that $K$ defines not only the discriminant group $\D$, but also a quadratic form on $\D$ with values in $\QQ/\ZZ$. This form is given by
\begin{equation}\label{def:quform}
\sq(X)=\frac{1}{2}K^{-1}({\tilde X},{\tilde X}), \quad X\in\D,
\end{equation}
where $\tilde X$ is an arbitrary lift of $X$ to $\Lambda^*$. It is easy to see that the fractional part of the right-hand-side of (\ref{def:quform}) is independent of the choice of the lift, and therefore $\sq$ is well-defined as a map from $\D$ to $\QQ/\ZZ$. This quadratic form gives rise to a $\QQ/\ZZ$-valued symmetric bilinear form on $\D$:
$$
\sb(X,Y)=\sq(X+Y)-\sq(X)-\sq(Y)=K^{-1}({\tilde X},{\tilde Y}).
$$
Since both the associator $h(A,B,C)$ and the braiding $s(A,B)$ are written in terms of $\sb$ and $\sq$, they only depend on $\D$, $\sq$, and the ordering of the generators of $\D$. One can check that changing the ordering modifies $h$ and $s$ by coboundary terms and therefore leads to an equivalent braided monoidal structure. Physically, this is rather obvious, since changing the order of the generators merely redefines the tensor product of Wilson lines by some Aharonov-Bohm phases.

\section{Topological boundary conditions}\label{sec:boundaries}

\subsection{The boundary gauge group}

Let $M_3$ be an oriented 3-manifold with a nonempty boundary $\Sigma=\partial M_3$. Consider an arbitrary variation of the gauge field $A$. The corresponding variation of the action contains a boundary term
$$
\delta_{bdry} S_{CS}=\frac{i}{4\pi}\int_\Sigma K(A_\para,\delta A_\para).
$$
Here and below the subscript $\para$ denotes the restriction of the 1-form to the boundary $\Sigma$.

The boundary term in the variation of the action defines a 1-form on the space of gauge fields on $\Sigma$. The differential of this 1-form is a symplectic form on the space of gauge fields. A consistent boundary condition should be local and define a Lagrangian submanifold with respect to this symplectic form. A topological boundary condition should also be invariant with respect to orientation-preserving diffeomorphisms of $\Sigma$.

Explicitly, the symplectic form is
$$\omega={1\over 4\pi}\int_{\Sigma}K(\delta A_{\para}, \delta A_{\para}).$$
A local diffeomorphism-invariant constraint on the $\g_\Lambda$-valued 1-form $A_\para$ requires it to lie in some subspace of $\g_\Lambda$. The corresponding submanifold is Lagrangian if and only if this subspace is isotropic with respect to the symmetric bilinear form $K$ and its dimension  is half the dimension of $\g_\Lambda$. Such a subspace exists if and only if the signature of $K$ vanishes. For such $K$ the dimension of isotropic subspaces ranges from $0$ to $\frac12 \dim \g_\Lambda$, so we will call an isotropic subspace of dimension $\frac12 \dim \g_\Lambda$ a Lagrangian subspace.

Note that the well-known boundary condition for  Chern-Simons theory leading to a chiral boson on the boundary is $A_\bz=0$, where $z$ is a local complex coordinate on $\Sigma$. This condition also defines a Lagrangian submanifold in the space of $A_\para$, but it depends on a choice of complex structure on $\Sigma$ and is not a topological boundary condition.

The condition that $A_\para$ lies in a Lagrangian subspace of $\g_\Lambda$ means that the gauge group on the boundary is broken down to a Lagrangian subgroup. Such a subgroup of $\TT_\Lambda$ may be connected or disconnected. If it is connected, then it is a torus $\TT_0$. If it is disconnected, then it is isomorphic to a product of a finite abelian group and a Lagrangian torus.

We will first study the case of a connected boundary gauge group. The case when the boundary gauge group is a product of a torus and a finite abelian group is more complicated and is considered in section 6.

A torus is a divisible abelian group and therefore we can always decompose $\TT_{\Lambda}$ into a product of $\TT_0$ and another torus. Thus if we denote $\Lambda_0=H_1(\TT_0,\ZZ)$, then $\Lambda_0$ is a subgroup of $\Lambda$, and moreover $\Lambda_0$ is a direct summand of $\Lambda$. That is, the exact sequence of abelian groups
$$
0\ra \Lambda_0\ra \Lambda\ra \Lambda/{\Lambda_0}\ra 0.
$$
splits. We will denote by $P$ the injective map from $\Lambda_0$ to $\Lambda$; if we choose generators for $\Lambda_0$ and $\Lambda$, $P$ is represented by an integral matrix of size $2n\times n$, where $2n$ is the rank of $\Lambda$.  The transposed matrix $P^t$ can be regarded as a surjective map from $\Lambda^*$ to $\Lambda_0^*$. The fact that $\TT_{\Lambda_0}$ is isotropic is equivalent to the matrix identity
$$
P^t K P=0.
$$

\subsection{Gauge-fixing in the presence of a boundary}

To proceed further we need to fix a gauge. We introduce a Lagrange multiplier field
$B$ taking values in the dual $\g_\Lambda^*$ of the Lie algebra $\g_\Lambda=\Lambda\otimes\RR$ of $\TT_\Lambda$,
as well as the Faddeev-Popov ghost field $C$ and the anti-ghost field ${\bar C}$ taking values in $\g_\Lambda$ and $\g_\Lambda^*$ respectively. The BRST
operator $\mathbf{Q}$ acts as \be\label{brst}\delta_{\mathbf{Q}} A=dC,
\quad \delta_{\mathbf{Q}}
C=0,\quad \delta_{\mathbf{Q}} {\bar C}=B,\quad \delta_{\mathbf{Q}} B=0.\ee

We will use the Lorenz gauge $d\star A=0$, where $\star $ is the Hodge star operator with respect to some Riemannian metric on $M_3$.
On a closed 3-manifold $M_3,$ the gauge-fixed
action has the form
\be \label{fullaction}
S_{CS+GF}=S_{CS}+\int_{M_3}\delta_{\mathbf{Q}}\Bigl(({\bar C},d\star A)\Bigr)=\int_{M_3}\Biggl({i\over 4\pi}K(A,dA)+(d{\bar C},\star dC)+(B, d\star A)\Biggr).\ee
Here $(\cdot,\cdot)$ denotes the pairing between an element of $\g_\Lambda^*$ and $\g_\Lambda$. This action is obviously BRST-invariant.

Now suppose $M_3$ has a boundary $\partial M_3=\Sigma$. Since the boundary gauge group is a subgroup of $\TT_\Lambda$, the boundary value of the field $C$ is restricted accordingly: it takes values in the subspace $\g_{\Lambda_0}$. To determine the rest of boundary conditions, we consider  the boundary term in the variation of the gauge-fixed action:
$$\delta_{bdry} S_{CS+GF}=
\frac{i}{2\pi} \int_{\Sigma}K\left(A_\para, \delta A_\para\right) +\int_{\Sigma}vol_{\Sigma}\Bigl(
(B,\delta A_{\perp})+(\delta {\bar C},\partial_\perp C) +(\partial_\perp {\bar C},\delta C)\Bigr).$$
Here and below the subscript $\perp$ denotes the component of a 1-form orthogonal to the boundary.

Requiring the vanishing of these boundary terms tell us first of all that $\partial_\perp {\bar C}$ takes values in the subspace of $\g_\Lambda^*$ which annihilates $\g_{\Lambda_0}$, i.e. in $\g_{\Lambda/\Lambda_0}^*$. The boundary value of ${\bar C}$ must lie in a complementary subspace of $\g^*_\Lambda$ which we denote $H$, and then $\partial_\perp C$ must lie in the annihilator of $H$. Finally, BRST-invariance requires $B$ to lie in $H$, and then $A_\perp$ must lie in in the annihilator of $H$.

A choice of a decomposition $\g^*_\Lambda\simeq \g_{\Lambda/\Lambda_0}^*\oplus H$ may be regarded as a choice of a splitting of an exact sequence of vector spaces
\begin{equation}\label{splittingvect}
0\ra  \g_{\Lambda/\Lambda_0}^*\ra \g_\Lambda^*\ra \g_{\Lambda_0}^*\ra 0.
\end{equation}
That is, we may describe $H$ as the image of an injective map
$$
R: \g_{\Lambda_0}^*\ra \g_\Lambda^*
$$
such that $P^tR=\Id$. Note that $R$ is a map of vector spaces and does not necessarily arise from a splitting of the exact sequence of the abelian groups
$$
0\ra (\Lambda/\Lambda_0)^*\ra\Lambda^*\ra\Lambda_0^*\ra 0.
$$
As discussed above, this sequence always splits, so we could, if we wished, choose $R$ to be a splitting of this exact sequence of free abelian groups. That is, we could choose $R$ to be an integral solution of the matrix equation $P^tR=\Id$. However, there is no reason to do this, so for the time being $R$ will remain a map of vector spaces. We will see later that there are some natural integrality constraints on $R$, but these constraints do not reduce to the statement that the matrix representing $R$ is integral.

To write all these boundary conditions in a compact form, let us define the projector $\Y$ from $\g_\Lambda$ to $H^*$ and the complementary projector $\Id-\Y$:
$$
\Y=\Id-PR^t,\quad \Id-\Y=PR^t.
$$
Then
$$
\ker \Y=\g_{\Lambda_0}, \quad \ker (\Id-\Y)=H^*,\quad \ker \Y^t=H,\quad \ker (\Id-\Y^t)=\g_{\Lambda/\Lambda_0}^*.
$$
These equalities are equivalent to the matrix identities
$$
\Y P=0,\quad R^t \Y=0.
$$

 The boundary conditions can now be written as follows:
 \be \label{condi}
\Y A_{\parallel}=0,\quad (\Id-\Y)
A_{\perp}=0,\quad \Y^t B=0,\quad
\Y C=0,\quad \Y^t{\bar C}=0\ee \be
\label{cond_ii} (\Id-\Y)\p_{\perp}C=0,\quad
(\Id-\Y)^t\p_{\perp}{\bar C}=0  \ee

Note that to prove BRST-invariance of these boundary
conditions we have to use the equation of motion for $A_{\perp}$ which
implies $\p_{\perp} B\vert_{\Sigma}\sim K dA_{\parallel}.$ Indeed, the BRST variation of the second equation in  (\ref{cond_ii})
gives:
$$(\Id-\Y)^t\p_{\perp}B\vert_{\Sigma}=0 $$
which is true due to $P^tKP=0.$

\subsection{Constraints on the splitting}

We will now discuss some natural constraints on the splitting $R$
which arise from considering quantization of charges in the presence
of a boundary. Let the space-time be a half-space $x^3>0$, with the
boundary located at $x^3=0$. Consider an electric charge
$X\in\Lambda^*$ whose worldline is given by $x^2=0,x^3=L$. We will
regard $x^1$ as time. We would like to solve for the static field
created by this charge using the method of images.  Let $\hX$ be the
image charge. Consider a point $\mathbf{P_*}$ on the boundary with
spatial coordinates $x^2=a, x^3=0$, let $\rho={\sqrt { L^2+a^2}}$ be
the distance from this point to the charge, and let $\alpha$ be the
polar angle of the charge measured from the point $\mathbf{P_*}$.

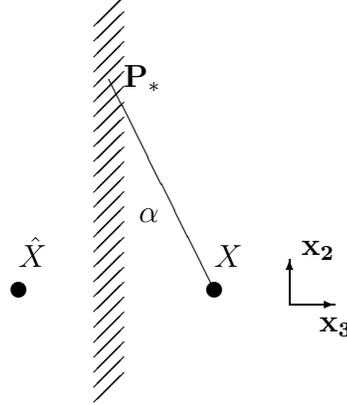
\setlength{\unitlength}{2mm}
\centerline{\begin{picture}(20,30)
  \linethickness{1.5mm}
  \multiput(14, 4)(0,1){26}{\line(1, 1){2}}
  \put(22,13){$X$}
  \put(22,11.5){\circle*{1}}
  \put(16,25){$\mathbf{P_*}$}
  \put(9,11.5){\circle*{1}}
  \put(9,13){$\hX$}
  \put(22,11.5){\line(-1,2){7}}
  \put(17,16){$\alpha$}
 % \qbezier(16,14)(16, 14.5)(17,15)
 \thinlines
\put(27, 10.5){\vector(1, 0){3}}
  \put(27, 10.5){\vector(0, 1){3}}
  \put(27.8, 13.8){$\mathbf{x_2}$}
  \put(29, 8.7){$\mathbf{x_3}$}
\end{picture}}

Then the gauge field at $\mathbf{P_*}$ is given by
 \be
\label{nearbi}A_3=K^{-1}(X+\hX){\cos \a\over \rho},\quad
 A_2=K^{-1}(\hX-X){\sin \a\over \rho}\ee
% Note in particular that when $q$ and $\hat q$ approach the boundary, we find $\alpha=0$ so that
% Wilson line at the boundary creates gauge field perpendicular to the boundary.
%%%%%%%%%%%%%%%%%%%%%%%%%%%%%%%%%%%%%%%%%%%%%%%%%%%%%%%%%%%%%%%%%%%%%%%%%%%%%%%%%%
%\begin{figure}
%\begin{center}
%\epsfig{file=./cs3.eps,width=.7\textwidth} \caption{The gauge field
%on the boundary is a sum of $A$ and $\hat A$ created by bulk charge
%$q$ and its image $\hat q$} \label{fig:cs3}
%\end{center}
%\end{figure}
%%%%%%%%%%%%%%%%%%%%%%%%%%%%%%%%%%%%%%%%%%%%%%%%%%%%%%%%%%%%%%%%%%%%%%%%%
Imposing the boundary conditions (\ref{condi}) leads to
$$(\Id-\Y)K^{-1}(X+\hX)=0,\quad \Y K^{-1}(\hX-X)=0$$ The sum of these two equations gives the image
charge: \be\label{image_charge}{ \hX}=K(2\Y-\Id)K^{-1}X.\ee

It seems natural to require the image charge to be integral. Indeed, instead of considering a Wilson line parallel to the boundary, we may consider a Wilson line piercing the boundary. Then the image Wilson line will also pierce the boundary and we would need to require its charge to be integral.  Therefore the following matrix must be integral:
\begin{equation}\label{integrality1}
2K\Y K^{-1}\in\Mat(2n,\ZZ).
\end{equation}

A further integrality constraint emerges if we consider a monopole operator of charge $m\in\Lambda$ on which a Wilson line of charge $Km$ terminates. This Wilson line is isomorphic to the trivial one, therefore the image Wilson line should also be isomorphic to the trivial one. That is, the electric charge of the image Wilson line must be of the form $K{\hat m}$ for some ${\hat m}\in\Lambda$. This implies
\begin{equation}\label{integrality2}
2\Y \in\Mat(2n,\ZZ).
\end{equation}

Thus the compatibility of the method of images with the quantization
of charges seems to require that the constraints
(\ref{integrality1},\ref{integrality2}) be satisfied. Note that for
a given Lagrangian subgroup $\Lambda_0\subset\Lambda$ it is
not obvious that a splitting $R$ exists such that both constraints
are satisfied. For example, we can always choose $R$ to be integral;
then $\Y$ is also integral, thereby satisfying (\ref{integrality2}).
However, the constraint (\ref{integrality1}) is still nontrivial and
may rule out some subgroups $\Lambda_0$ which would otherwise be
allowed.

We will see later that the physical properties of boundary conditions are actually independent  of $\Y$ and all formulas make perfect sense for an arbitrary splitting $\Y$ not satisfying any integrality constraints. We therefore believe that the above integrality constraints on $\Y$ are an artifact of the method of images, and probably
there is a better approach which avoids them altogether.

\subsection{Examples of topological boundary conditions}

Let us describe a concrete example of a topological boundary condition in an abelian Chern-Simons theory.
We will make use of it in sections 5 and 6. Let the gauge group be $\TT_\Lambda=U(1)^{2m}$ and the coupling matrix
$K$ be block-diagonal
\begin{equation}\label{CSexample1}
K=\begin{pmatrix} \mathbb{K}_m & 0\\ 0 & -\mathbb{K}_m\end{pmatrix},\quad \mathbb{K}_m\in Mat(m,\ZZ).
\end{equation}
An obvious choice of a connected Lagrangian subgroup in $\TT_{\Lambda}$ is the torus corresponding to the sublattice $\Lambda_0\subset \Lambda$ consisting of elements of the form
$$
\begin{pmatrix} \lambda \\ \lambda\end{pmatrix},\quad \lambda\in\Lambda_0.
$$
The corresponding matrix $P$ has the form
$$
P=\begin{pmatrix} \Id_m \\ \Id_m\end{pmatrix}
$$
where $\Id_m$ is the $m\times m$ identity matrix.

The most general splitting $R$ satisfying $P^tR=\Id_m$ is given by
\begin{equation}\label{splittingex1}
R=\frac12 \begin{pmatrix} \Id_m+r \\ \Id_m-r\end{pmatrix}
\end{equation}
where $r$ is an $m\times m$ real matrix. This implies
$$2\Y=\begin{pmatrix} \Id_m-r^t & -(\Id_m-r^t)\\ -(\Id_m+r^t) & \Id_m+r^t\end{pmatrix}, \quad
2K\Y K^{-1}=\begin{pmatrix} \Id_m-\mathbb{K}_m r^t \mathbb{K}_m^{-1} & \Id_m-\mathbb{K}_m r^t \mathbb{K}_m^{-1}\\
\Id_m+\mathbb{K}_m r^t \mathbb{K}_m^{-1}& \Id_m+\mathbb{K}_m r^t
\mathbb{K}_m^{-1}\end{pmatrix}.$$ The two integrality constraints
(\ref{integrality1}) and (\ref{integrality2}) in this case are:
$$r \in Mat(m,\ZZ),\quad \mathbb{K}_m^{-1}r\mathbb{K}_m\in Mat(m,\ZZ).$$
The discriminant group is a direct sum
$$
\D=\bigoplus_{\c=1}^n \ZZ_{M_\c} \oplus
\bigoplus_{\c=1}^n \ZZ_{M_\c}
$$
where $M_\c$ are determined from
$\mathbb{K}_m$. In the simplest special case $\mathbb{K}_m=2N \Id_m$
with $N \in \ZZ$ the lattice $\Lambda$ is isomorphic to
$\ZZ^{2m}$, the sublattice $\Lambda_0$ is isomorphic to $\ZZ^{m},$ and the
discriminant group is $\D\simeq \ZZ_{2N}^{\oplus 2m}.$

In section 6.2 we will consider an example of a boundary condition in a Chern-Simons theory with a non-block-diagonal $K$.

\section{Boundary line operators}\label{sec:bdrylines}

\subsection{Boundary Wilson lines}

Line operators on a particular boundary are objects of a monoidal category. Morphisms in this category arise from local operators sitting at the joining point of two line operators, while the monoidal structure arises from the fusion of line operators. Fusion is associative, in a suitable sense, but is not commutative, in general. The most obvious boundary line operators are boundary Wilson lines and their direct sums. We will argue below that in fact there are no other boundary line operators.

Since the boundary gauge group $\TT_{\Lambda_0}$ is a subgroup of the bulk gauge group  $\TT_\Lambda$, the group of boundary charges $\Hom(\TT_{\Lambda_0},U(1))=\Lambda_0^*$ is a quotient of the group of bulk charges $\Hom(\TT_\Lambda,U(1))=\Lambda^*$. This also means that a boundary Wilson line can be regarded as a result of fusing  a bulk Wilson line with the boundary.

A bulk Wilson line can end on the boundary if its endpoint is not charged with respect to the boundary gauge group, or if its charge can be screened by a monopole operator.
This means that the charge of such a Wilson line must lie in the subgroup $(\Lambda/\Lambda_0)^*\oplus \Im\, K$ of $\Lambda^*$. If we identify Wilson lines which are isomorphic in the category of bulk line operators, we need to quotient this subgroup by $\Im\, K$. Thus, once we take screening by monopole operators into account, Wilson lines ending on the boundary have charges in the subgroup $\sM\subset\D$ where
$$
\sM=((\Lambda/\Lambda_0)^*\oplus \Im\, K)/\Im\, K.
$$

Monopole operators also affect the classification of boundary Wilson lines. Morphisms between different boundary Wilson lines arise from monopole operators which carry electric charge $K m$ for some $m\in\Lambda$. Here we should regard $Km$ not as an element of $\Lambda^*$, but as an element of its quotient $\Lambda_0^*$. In other words, two boundary Wilson lines are isomorphic if and only if the difference of their charges lies in the sublattice $\Im\, P^t K$ of $\Lambda_0^*$. (We follow the conventions of the previous section and denote by $P^t$ the surjective map from $\Lambda^*$ to $\Lambda_0^*$.)  Thus isomorphism classes of boundary Wilson lines are labeled by elements of the finite abelian group $\D_0=\Lambda_0^*/\Im\, P^t K$ which we will call the boundary discriminant group. We can write $\D_0$ a bit differently by noting that $P^tK$ annihilates $\Lambda_0$ (by virtue of $P^tKP=0$), and therefore can be regarded as a map $K_0$ from $\Lambda/\Lambda_0$ to $\Lambda_0^*$. Then $\D_0=\Lambda_0^*/\Im\, K_0$.

We have seen above that properties of bulk line operators are determined by the discriminant group $\D$ and the $\QQ/\ZZ$-valued quadratic form $\sq$ on it. We will argue in section 7 that for fixed signature these data determine the isomorphism class of the abelian Chern-Simons theory. Thus it should be possible to describe properties of boundary line operators in terms of $\D$ and $\sq$ (the signature is zero for topological boundary conditions to exist). To do this, note that the boundary discriminant group $\D_0$ can also be written as
$$
\D_0=\D/\sM.
$$
where $\sM$ is the subgroup of $\D$ classifying Wilson lines which can end on the boundary. Thus as far as the classification of boundary Wilson lines is concerned, the important object is the subgroup $\sM$. We will see later that the boundary associator also depends only on $\D,\sq,$ and $\sM$.

Recall that on the classical level the boundary condition is determined by a choice of a sublattice $\Lambda_0\subset\Lambda$ such that the subspace $\Lambda_0^\QQ=\Lambda_0\otimes\QQ$ is a Lagrangian subspace of $\Lambda^\QQ=\Lambda\otimes\QQ$. This means that the metric $K$ vanishes when restricted to $\Lambda_0^\QQ$ (i.e. $\Lambda_0^\QQ$ is isotropic), and that the orthogonal complement of $\Lambda_0^\QQ$ in $\Lambda^\QQ$ is contained in $\Lambda_0^\QQ$ (i.e. $\Lambda_0^\QQ$ is coisotropic).

On the quantum level the relevant data are the discriminant group $\D$ and the quadratic form $\sq$, while the boundary condition is described by a subgroup $\sM$. Let us show that $\sM$ is a Lagrangian subgroup of $\D$, in the sense that it is both isotropic and coisotropic with respect to the quadratic form $\sq$. To show that it is isotropic, note that an element $X\in \sM$ can be lifted to an element ${\tilde X}\in \Lambda^*$ which annihilates $\Lambda_0$ plus an element  in $\Im K$. The element in $\Im K$ does not affect the quadratic form $\sq$ and may be ignored. The element ${\tilde X}\in\Lambda^*$ is conormal to the subspace $\Lambda_0\otimes\QQ$ of $\Lambda\otimes\QQ$. Since by assumption $\Lambda_0\otimes\QQ$ is a Lagrangian subspace, ${\tilde X}$ can be written as $K\lambda$ for some $\lambda\in\Lambda_0\otimes\QQ$. But then
$$
\sq(X)=\frac12 K^{-1}({\tilde X},{\tilde X})=\frac12 K(\lambda,\lambda)=0.
$$
so $\sM$ is isotropic.

To show that $\sM$ is coisotropic, consider an element $Y\in\D$ such that  $\sb(X,Y)=0$ for all $X\in\sM$. This implies that once we lift $Y$ to ${\tilde Y}\in\Lambda^*$,  it will satisfy
$$
K^{-1}({\tilde X},{\tilde Y})\in\ZZ,\quad \forall {\tilde X}\in (\Lambda/\Lambda_0)^*.
$$
Now recall that $\Lambda$ splits into a direct sum of $\Lambda_0$ and $\Lambda/\Lambda_0$. Once we choose the splitting, we can decompose $K^{-1}{\tilde Y}$ into a component in $\Lambda_0\otimes\QQ$ and a component in $(\Lambda/\Lambda_0)\otimes\QQ$. The above condition says that the latter component actually belong to the lattice $(\Lambda/\Lambda_0)$ while the former component is unconstrained. Thus $K^{-1}{\tilde Y}$ lies in the set $(\Lambda/\Lambda_0)\oplus (\Lambda_0\otimes\QQ)$, and therefore
$$
{\tilde Y}=K\lambda+KPr,
$$
where $\lambda\in \Lambda/\Lambda_0$ and $r\in\Lambda_0\otimes \QQ$. The first term lies in $\Im\, K$, while the second term lies in $(\Lambda/\Lambda_0)^*$. Hence
$Y\in\sM$, which means that $\sM$ is coisotropic.

It is tempting to conjecture that one can associate a boundary condition to any Lagrangian subgroup of $\D$. Some further evidence for this will be presented in section \ref{sec:generalbdries}, where we will construct more general boundary conditions and will see that they also correspond to Lagrangian subgroups in $\D$.

Physical quantities should not change if one replaces a boundary Wilson line with another one which is isomorphic to it. Let us verify this for a simple observable which is analogous to the Hopf link in the bulk theory. Consider a boundary Wilson line with charge $z\in\Lambda_0^*$ located at $x^2=0$ and a bulk Wilson line with charge $X\in\Lambda^*$ whose shape is a semi-circle in the $(x^2,x^3)$ plane with the center at the origin.

\vspace{1cm}

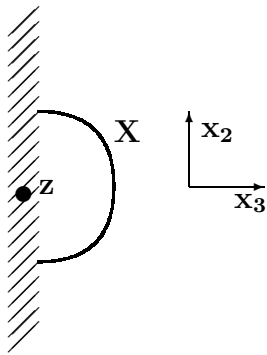
\setlength{\unitlength}{2mm}
\centerline{\begin{picture}(30, 20)
  \linethickness{1.5mm}
   \multiput(13, -1)(0,1){22}{\line(1, 1){2}}
   \put(15,9.5){${\mathbf z}$}
   \put(14,9.5){\circle*{1}}
  \thicklines
  \qbezier(20, 10)(20, 15)(15, 15)
  %\qbezier(3, 3)(2, 3)(2, 2)
  %\qbezier(2, 2)(2, 1)(3, 1)
  \qbezier(15, 5)(20, 5)(20, 10)
  \put(20,13){$\mathbf{X}$}
  \thinlines
  \put(25, 10){\vector(1, 0){5}}
  \put(25, 10){\vector(0, 1){5}}
  \put(25.8, 13.3){$\mathbf{x_2}$}
  \put(28, 8.6){$\mathbf{x_3}$}
\end{picture}}

\vspace{1cm}

Since both ends of the bulk Wilson line are on the boundary, $X$ must satisfy $P^t X=0$ (unless we insert monopole operators on the endpoints). We will call this configuration of Wilson lines a half-link. Its expectation value can be computed as follows. Lifting $z\in\Lambda_0^*$ to an element $Z\in\Lambda^*$, we may regard the boundary Wilson line as a limit of a bulk Wilson line with charge $Z$. When we fuse it with the boundary, it collides with its image charge $K(2\Y-1)K^{-1}Z$, so the field created by a boundary charge can be computed as the field of a bulk charge $2K\Y K^{-1}Z$:
$$
A=2\Y K^{-1}Z d\theta
$$
where $\theta$ is the angular coordinate in the $(x^2,x^3)$ plane. The expectation value of the half-link can be computed as the Aharonov-Bohm phase of the charge $X$ moving along the half-circle parameterized by $\theta\in[-\pi/2,\pi/2]$:
$$
\exp(-i\int X(A))=\exp(-2\pi i (X,\Y K^{-1}Z)).
$$
It follows from $P^t X=0$ that $\Y^t X=X$, so the above phase can also be written as
$$
\exp(-2\pi i K^{-1}(X,Z)).
$$
Note first of all that it is independent of the lift of $z$ to $Z$. Indeed, different lifts differ by elements of $\ker P^t$. Since $\g_{\Lambda_0}$ is Lagrangian, any element in $\ker P^t$ can be written as $KPf$ for some $f\in \g_{\Lambda_0}$. But then the Aharonov-Bohm phases differ by a factor
$$
\exp(-2\pi i K^{-1}(X,KPf))=\exp(-2\pi i (P^t X,f))=1.
$$
Second, if we shift either $X$ or $Z$ by an element of of the form $Km$, $m\in \Lambda$, the Aharonov-Bohm phase also does not change. This shows that replacing a Wilson line (either bulk or boundary) with an isomorphic one does not change the expectation value of the half-link. We can make it explicit if we regard $X$ as taking values in $\sM=(\Lambda/\Lambda_0)^*/\Im K_0^t$ and $z$ as taking values in $\D_0=\Lambda_0^*/\Im K_0=\D/\sM$. Then the expectation value of the half-link is
$$
\exp(-2\pi i \sb(X,Z)),
$$
where $Z$ is a lift of $z$ to $\D$. The precise choice of the lift is unimportant, since the bilinear form $\sb:\D\times\D\ra\QQ/\ZZ$ vanishes when restricted to $\sM\times\sM$.

\subsection{Semi-braiding and the boundary associator}
Fusing two boundary Wilson lines with charges $x,y\in \D_0=\D/\sM$ gives
a boundary Wilson line with charge $x+y$.  In order to completely
specify the monoidal structure on the category of boundary line
operators we also need to determine the associator morphism.
% This
%boundary associator should be compatible with semi-braiding, a
%certain analog of braiding which arises if we consider the
%relationship between bulk and boundary Wilson lines.
%We will do it in the next subsection.
 Note that while the
isomorphism class of the resulting Wilson line does not change if we
exchange $x$ and $y$, there is no natural isomorphism between
$W_x\otimes W_y$ and $W_y\otimes W_x$. That is, the monoidal
category of boundary Wilson lines does not have a natural symmetric
or even braided structure.

On the other hand, a certain analog of braiding arises if we
consider the relationship between bulk and boundary Wilson lines.
Consider a bulk Wilson line with charge $X\in\D$. Fusing it with the boundary gives a boundary Wilson
line whose charge (an element of $\D/\sM$) is the image of $X$ under the natural projection $\pi:\D\ra\D/\sM$. We will denote this charge $\pi(X)$. Given a boundary Wilson line
with charge $y\in\D/\sM$ we may consider fusing $W_y$ with $W_{\pi(X)}$ in two
different orders. In this case there is a natural isomorphism
between $W_y\otimes W_{\pi(X)}$ and $W_{\pi(X)}\otimes W_y$ arising
from the fact that one can change the order of boundary Wilson lines
by moving one of them to the bulk. Of course, this natural
isomorphism depends on $X\in\D$, not just its image $\pi(X)$
in $\D/\sM$. We will call this natural isomorphism a
semi-braiding between the bulk Wilson line $W_X$ and the boundary
Wilson line $W_y$.

Let us explain the mathematical meaning of the semi-braiding.
In the theory of monoidal categories there is a notion of a Drinfeld
center. It is defined by analogy with the center of an associative
algebra. The Drinfeld center of a monoidal category $\cC$ is a
braided monoidal category $\Z(\cC)$ whose objects are pairs
$(a,\chi_a)$, where $a$ is an object of $\cC$ and $\chi_a$ is a family
of isomorphisms $b\otimes a\ra a\otimes b$ for all objects
$b$ of $\cC$. These isomorphisms must satisfy some compatibility constraints with the tensor product in the category $\cC$, see appendix B for details. It turns out that the semi-braiding defines  a map from the set of bulk line operators to the set of objects of the
Drinfeld center of the category of boundary line operators. In fact, it can be shown that the map on objects can be extended to a strong monoidal functor from
the category of bulk line operators to the Drinfeld
center of the category of boundary line operators. An interested reader is referred to
appendix B for mathematical details.

Let us compute the semi-braiding isomorphism between the bulk Wilson
line $W_X, X\in \D$ and
 a boundary Wilson line $W_y, y\in \D_0$.
 %Let us first assume that both bulk and boundary discriminants
  %are simple $\D=Z_M$ and $\D_0=Z_m$.
  To write down a concrete formula, one needs to choose isomorphisms $W_y\otimes W_{y'}\simeq W_{y+y'}$
  for all boundary Wilson lines, so that the semi-braiding becomes an isomorphism from
  $W_{\pi(X)+y}$ to itself,
   i.e. a complex number. An important ingredient in the definition of this isomorphism  and in the
   further construction of the semi-braiding is the Aharonov-Bohm
phase arising from the counterclockwise transport
of charge $X$ along a semi-circle centered at the boundary charge $y$. (We
have arranged Wilson lines so that both $W_X$ and $W_y$ are along $x^1$ direction
and $W_y$ sits at $x^2=x^3=0$). The computation of this phase is
similar to the computation of the expectation value of the
half-link, except that the bulk charge $X$ is not assumed to be
annihilated by $P^t$.
 We use the method of images to replace the computation in the presence of a boundary with the computation
  without the boundary but with image charges included.
\vspace{1cm}

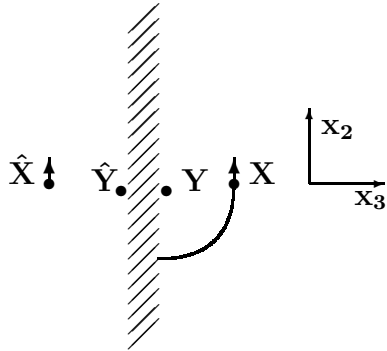
\setlength{\unitlength}{2mm} \centerline{
\begin{picture}(30, 20)
  \linethickness{1.5mm}
   \multiput(13, -1)(0,1){22}{\line(1, 1){2}}
   \put(16.5,9.5){$\mathbf{ Y}$}
   \put(15.5,9.5){\circle*{0.7}}
\put(10.5,9.5){$\mathbf{\hat {Y}}$}
   \put(12.5,9.5){\circle*{0.7}}
\thicklines
 % \qbezier(20, 10)(20, 15)(15, 15)
  %\qbezier(3, 3)(2, 3)(2, 2)
  %\qbezier(2, 2)(2, 1)(3, 1)
  \qbezier(15, 5)(20, 5)(20, 10)
  %\qbezier(10, 5)(15, 5)(10, 10)
  \put(21,10){$\mathbf{X}$}
  \put(20,10){\circle*{0.7}}
   \put(20, 10){\vector(0, 1){1.75}}
   \put(5,10){$\mathbf{\hX}$}
  \put(7.7,10){\circle*{0.7}}
   \put(7.7, 10){\vector(0, 1){1.75}}
  \thinlines
  \put(25, 10){\vector(1, 0){5}}
  \put(25, 10){\vector(0, 1){5}}
  \put(25.8, 13.3){$\mathbf{x_2}$}
  \put(28, 8.6){$\mathbf{x_3}$}
\end{picture}}

\vspace{1cm}

The boundary charge $y\in\Lambda_0^*$ is thus lifted to a bulk charge $Y\in\Lambda^*$ and moved slightly away from the boundary. Its image charge is $\hY=K(2\Y-1)K^{-1}Y$. The image charge for the bulk charge $X$ is $\hX=K(2\Y-1)K^{-1}X$. The Aharonov-Bohm phase is a sum of two terms: the phase acquired by the charge $X$ while moving counterclockwise in the static field created by $Y+\hY$ and the phase acquired by the charge $Y$ due to a time-like gauge field created by moving charges $X$ and $\hX$. The latter phase is the same as the phase that the charge $Y$ would acquire if it moved counterclockwise in the static field created by the charge $X-\hX$. The first phase is
$$
\exp\left(-\frac{i\pi}{2}K^{-1}(X,Y+\hY)\right).
$$
The second phase is
$$
\exp\left(-\frac{i\pi}{2}K^{-1}(X-\hX,Y)\right).
$$
Their product is
$$
\exp\left(-\pi i (X,\sbh Y)\right),\quad \sbh=K^{-1}+\Y K^{-1}-K^{-1}\Y^t.
$$
Note that this Aharonov-Bohm phase is independent of the choice of
lifting $y\in\Lambda_0^*$ to $Y\in\Lambda^*$. Indeed, any two such
lifts differ by an element of the form $KPr$ for some $r\in
\g_{\Lambda_0}$. But $\Y^tKPr=KPr$, and $\Y K^{-1}KPr=0$, so the phase
is unchanged if we shift $Y$ by $KPr$.

Now we are ready to define the product of boundary Wilson lines.
Let bulk and boundary discriminant groups be $\D=\prod_{i=1}^N
\ZZ_{M_i}$ and $\D_0=\prod_{\c=1}^n \ZZ_{m_\c}$ respectively. Let
$v_1< v_2 \ldots < v_n$ be an ordered set of generators of $\D_0.$
For all $\c\in\{1,\ldots,n\}$ we choose a lift of $v_\c \in \D_0$ to $e_\c\in
\D$ . We also choose generators $e_{n+1},\ldots,e_N$ for the subgroup $\sM$. We therefore get an ordered set of generators $e_1<\ldots<e_N$ for $\D$.
In order to define the product of boundary Wilson lines we need to lift $e_i$ to $\Lambda^*$. We denote these lifts $\te_i$, $i=1,\ldots,N$ and write
$$
\tv_\c=\te_\c,\quad \c=1,\ldots,n.
$$
We can always choose the lift of the remaining $N-n$ generators so that $\te_{n+1},\ldots,\te_N$ lie in the Lagrangian sublattice $(\Lambda/\Lambda_0)^*$, i.e.  there exist $r_{\hat i}\in\Lambda_0\otimes\QQ$, ${\hat i}\in\{n+1,\ldots,N\}$ such that
\be\label{choice}
{\tilde e}_{\hat j}=KPr_{\hat j} \quad {\hat j}=n+1,\ldots, N.
\ee
Note also that since by assumption $v_\gamma$ generates $\ZZ_{m_\gamma}$, we have $m_\gamma v_\gamma=0$ in $\D_0$, and therefore $m_\gamma e_\gamma\in \sM$, and therefore $m_\gamma \te_\gamma\in (\Lambda/\Lambda_0)^*\oplus\Im K$. Explicitly, this means that there exist $f_\gamma\in\Lambda_0\otimes\QQ$ such that
$$
m_\gamma \te_\gamma-KPf_\gamma\in\Im K,\quad \gamma=1,\ldots,n.
$$
We also have
$$
M_i \te_i\in\Im K,\quad i=1,\ldots,N.
$$

Recall (see section 3) that we think of a bulk Wilson line $W_{\vec
A}$ with charge ${\vec A}=A_1e_1+\ldots+A_Ne_N$ as the ordered
product of bulk Wilson lines $W_{A_1e_1},\ldots,W_{A_Ne_N}$. Similarly, we
define a boundary Wilson line $W_{\vec a}$ with charge ${\vec a}=a_1v_1+\ldots
+a_nv_n$ as the ordered product of boundary Wilson lines
$W_{a_1v_1},\ldots, W_{a_nv_n}$. For brevity we will refer to
$W_{a_i v_i}$ as constituents of $W_{\vec a}$. When fusing $W_{\vec a}$
with $W_{\vec b}$ we first need to rearrange the constituents to put
them into the standard order. This is possible since we chose the
lift of $v_\c$ to $e_\c$ and hence we can lift the boundary Wilson line $W_{b_\c v_\c}$ to the bulk
Wilson line $W_{b_\c e_\c}$ and move it counterclockwise around $W_{a_\b v_\b}$ for $\c<\b$.
Second, in general the $\c$-th component of of $\vec a +\vec b$ is
outside the standard range $[0,m_\c-1]$. To bring it to the standard
range we write
$$a_\c+b_\c=(a_\c\plusd b_\c)+m_\c\left [{a_\c+b_\c\over m_\c}\right]$$
where brackets denote the integral part, and move the `trivial'
boundary  lines $I_{(a_\c,b_\c)}$ with charges
$$v_\c m_\c\left [{a_\c+b_\c\over m_\c}\right]$$
to the left of all other constituent boundary lines.
The net result is a boundary Wilson line $W_{\vec a \plusd \vec
b}$ together with the trivial  boundary Wilson line $I_{(\vec a,\vec b)}$ with
charge
$$\sum_{\c=1}^n v_\c m_\c\left [{a_\c+b_\c\over m_\c}\right]$$
to the left of $W_{\vec a \plusd \vec b}.$

In going from $W_{\vec a}\, W_{\vec b}$ to $ W_{\vec a \plusd
\vec b}$ we get a phase given by (in additive notation)
$${\hat \psi}(\vec a,\vec b)=-\half \sum_{\a<\c} a_\c b_\a \sbh_{\a \c}-\half \sum_{\a>\c} m_\a\left [{a_\a+b_\a\over
m_\a}\right](a_\c\plusd b_\c)\sbh_{\a \c}$$ where $\sbh_{\a
\c}=(\tv_\a,\sbh \tv_\c).$ This is illustrated for $n=2$ below.

  \vspace{1cm}

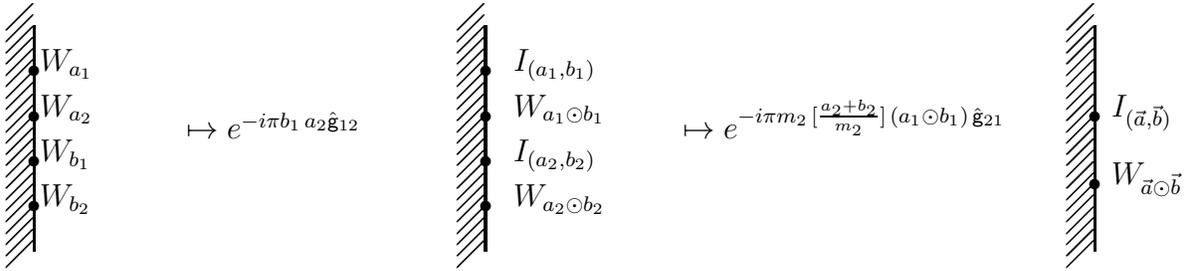
\setlength{\unitlength}{3cm}
\begin{picture}(1,1)
 \linethickness{0.3mm}
\put(0.125, 0){\line(0, 1){1}} \multiput(0,
-0.07)(0,0.05){22}{\line(1,1){0.12}} \put(0.125,
0.2){\circle*{0.05}}\put(0.15, 0.2){$W_{b_2}$}

\put(0.125, 0.4){\circle*{0.05}}\put(0.15, 0.4){$W_{b_1}$}

\put(0.125, 0.6){\circle*{0.05}}\put(0.15, 0.6){$W_{a_2}$}

\put(0.125, 0.8){\circle*{0.05}} \put(0.15, 0.8){$W_{a_1}$}

\put(0.8, 0.5){$\mapsto e^{-i\pi b_1 \, a_2 \sbh_{12}}$}

\put(2.125, 0){\line(0, 1){1}}
\multiput(2,-0.07)(0,0.05){22}{\line(1,1){0.12}}

 \put(2.125,
0.2){\circle*{0.05}}\put(2.25, 0.2){$W_{a_2\plusd b_2}$}

\put(2.125, 0.4){\circle*{0.05}}\put(2.25, 0.4){$I_{(a_2,b_2)}$}

\put(2.125, 0.6){\circle*{0.05}}\put(2.25, 0.6){$W_{a_1\plusd b_1}$}

\put(2.125, 0.8){\circle*{0.05}} \put(2.25, 0.8){$I_{(a_1,b_1)}$}

\put(3, 0.5){$\mapsto e^{-i\pi m_2 \, [{a_2+b_2\over
m_2}]\,(a_1\plusd b_1)\, \sbh_{21}}$}

\put(4.825, 0){\line(0, 1){1}}\multiput(4.7,
-0.07)(0,0.05){22}{\line(1,1){0.12}} \put(4.825,
0.3){\circle*{0.05}} \put(4.9,0.3){$W_{\vec a\plusd \vec b}$}

\put(4.825, 0.6){\circle*{0.05}} \put(4.9,0.6){$I_{(\vec a,\vec
b)}$}
 \end{picture}

 \vspace{1cm}

 Note also that fusing a bulk Wilson line $W_{\vec B}$ with the boundary
 gives a product of boundary Wilson lines $I_{\vec B} \, W_{\pi(\vec B)}$ where
 $I_{B_\a}$ is a trivial Wilson line with charge $v_\a m_\a \tB_\a$ and we denote
 $$\pi({\vec B})_\a=B_\a-m_\a\tB_\a,\quad  \tB_\a=\left[B_\a\over m_\a \right].$$

  Now we are ready to compute the semi-braiding isomorphism
$${\mathbf c}^{(sb)}_{\vec a \vec B}: W_{\vec a\plusd \pi(\vec B)}\mapsto W_{\pi(\vec B)\plusd
\vec a}$$
 which is defined (see cartoon below) by inverting the first and the second arrows and  following
the other three in the direction indicated.

\vspace{1cm}

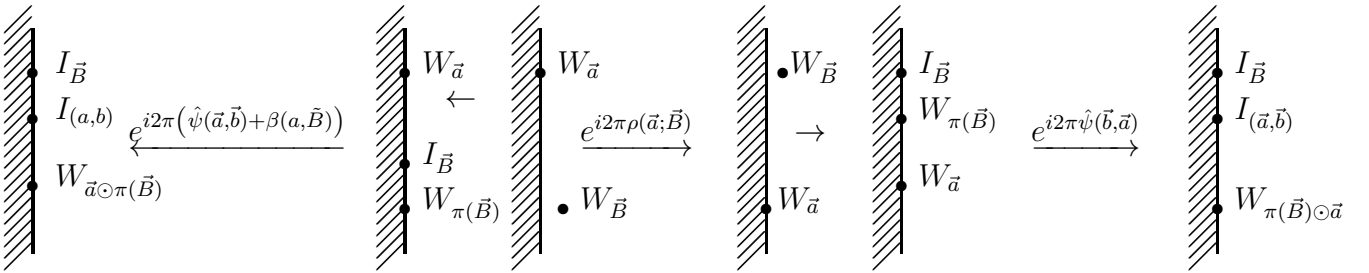
\setlength{\unitlength}{3cm}
\begin{picture}(1,1)
 \linethickness{0.3mm}

\put(-0.125, 0){\line(0, 1){1}} \multiput(-0.25,
-0.07)(0,0.05){22}{\line(1,1){0.12}}

\put(-0.125, 0.8){\circle*{0.05}}\put(-0.025, 0.8){$I_{\vec B}$}
\put(-0.125, 0.6){\circle*{0.05}}\put(-0.025, 0.6){$I_{(a,b)}$}
 \put(-0.125,
0.3){\circle*{0.05}}\put(-0.025, 0.3){$W_{\vec a \plusd \pi(\vec
B)}$}

\put(0.3, 0.5){$\underleftarrow{ e^{i2\pi\left( {\hat \psi}(\vec a,
\vec b)+\beta(a,\tB)\right)}}$}

\put(1.525, 0){\line(0, 1){1}}
\multiput(1.4,-0.07)(0,0.05){22}{\line(1,1){0.12}}

 \put(1.525,
0.2){\circle*{0.05}}\put(1.6, 0.2){$W_{\pi(\vec B)}$}

\put(1.525, 0.4){\circle*{0.05}}\put(1.6, 0.4){$I_{\vec B}$}
%\put(1.625, 0.4){\circle*{0.05}}\put(1.7, 0.4){$W_{a_1}$}

\put(1.525, 0.8){\circle*{0.05}}\put(1.6, 0.8){$W_{\vec a}$}

\put(1.7, 0.65){$\leftarrow$}

\put(2.125, 0){\line(0, 1){1}}
\multiput(2,-0.07)(0,0.05){22}{\line(1,1){0.12}} \put(2.125,
0.8){\circle*{0.05}}\put(2.2, 0.8){$W_{\vec a}$}

\put(2.225, 0.2){\circle*{0.05}}\put(2.3, 0.2){$W_{\vec B}$}

%\put(1.625, 0.8){\circle*{0.05}} \put(1.7, 0.8){$I_{(\vec a,\vec
%b)}$}

\put(2.3, 0.5){$\underrightarrow{e^{i2\pi \rho(\vec a;\vec B)}}$}

\put(3.125, 0){\line(0, 1){1}}
\multiput(3,-0.07)(0,0.05){22}{\line(1,1){0.12}}

 \put(3.125,
0.2){\circle*{0.05}}\put(3.175, 0.2){$W_{\vec a}$} \put(3.2,
0.8){\circle*{0.05}}\put(3.23, 0.8){$W_{\vec B}$} \put(3.25,
0.5){$\rightarrow$}
%\put(3.325, 0.4){\circle*{0.05}}\put(3.375, 0.4){$W_{b_1}$}

%\put(3.325, 0.6){\circle*{0.05}}\put(3.375, 0.6){$W_{\pi(\vec B)}$}

\put(3.725, 0){\line(0, 1){1}} \multiput(3.6,
-0.07)(0,0.05){22}{\line(1,1){0.12}}

\put(3.725, 0.8){\circle*{0.05}}\put(3.8, 0.8){$I_{\vec B}$}
\put(3.725, 0.6){\circle*{0.05}}\put(3.8, 0.6){$W_{\pi(\vec B)}$}
 \put(3.725,
0.3){\circle*{0.05}}\put(3.8, 0.3){$W_{\vec a}$}

\put(4.3, 0.5){$\underrightarrow{ e^{i2\pi {\hat \psi}(\vec b, \vec
a)}}$}
%\put(3.325, 0.8){\circle*{0.05}} \put(3.375, 0.8){$W_{a_1}$}

\put(5.125, 0){\line(0, 1){1}}\multiput(5,
-0.07)(0,0.05){22}{\line(1,1){0.12}} \put(5.125,
0.2){\circle*{0.05}} \put(5.2,0.2){$W_{\pi(\vec B)\plusd \vec a}$}
\put(5.125, 0.6){\circle*{0.05}} \put(5.2,0.6){$I_{(\vec a,\vec
b)}$} \put(5.125, 0.8){\circle*{0.05}}\put(5.2,0.8){$I_{\vec B}$}
 \end{picture}

 \vspace{1cm}
 Thus the semi-braiding isomorphism is the phase ${\mathbf c}^{(sb)}_{\vec
 a \vec B}=e^{2\pi i s_{sb}(\vec a;\vec B)}$ where
 \be\label{semi_fin}s_{sb}(\vec a;\vec B)=
 s^{(0)}_{sb}(\vec a;\vec B)+{\hat \psi}(\vec b,\vec a)-{\hat \psi}(\vec a,\vec
 b),\quad s^{(0)}_{sb}(\vec a;\vec B)=\rho(\vec a;\vec B) -\beta(\vec a,\vec \tB),\quad \vec b=\pi(\vec B).\ee
The terms with $\hat \psi$ arise from our definition of the tensor products of
boundary Wilson lines. On the other hand $\rho(\vec a;\vec B)$ is a phase which arises
when the bulk line $W_{\vec B}$ is moved counterclockwise to the
left of the boundary Wilson line $W_{\vec a}$:
$$\rho(\vec a;\vec B)=-\half \sum_{j=1}^N \sum_{\c=1}^n B_j a_\c \left(\tilde e_j,\sbh\tv_\c\right)$$
Finally, $\beta(a,\tB)$ is a phase which arises in going from $W_{\vec a}I_{\vec B}
W_{\pi(\vec B)}$ to $I_{\vec B}W_{\vec a}W_{\pi(\vec B)}$ by moving
$W_{\vec a}$ clockwise to the right of $I_{\vec B}$:
$$\beta(a;\tB)=\half \sum_{\a,\c} a_{\c} m_{\a}\tB_{\a}\sbh_{\c
\a}.$$

We can further simplify this using (\ref{choice}) and $$(KPr_{\hat
j},\sbh\tv_\c)=2(r_{\hat j},P^t\tv_\c) .$$
The final answer for the semi-braiding is \be
\label{semi_finii}s_{sb}(\vec a;\vec B)=-\half \sum_{\c=1}^n b_\c
a_\c(\tv_\c, K^{-1}\tv_\c) -\sum_{\a > \c}b_\a a_\c (\tv_\a,
K^{-1}\tv_\c)\ee $$-\sum_{\hat j=n+1}^N\sum_{\c=1}^n B_{\hat j} a_\c
(r_{\hat j}, P^t\tv_\c)- \sum_{\a,\c} a_{\c}
m_{\a}\tB_{\a}(\tv_\c,K^{-1}\tv_\a).$$
Note that the dependence on the projector $\Y$ dropped out of the final expression.

Now let us compute the boundary associator:
$${\mathbf a}^{(bry)}_{a,b,c}:\left(W_{\vec{a}}\otimes W_{\vec{b}}\right)\otimes
W_{\vec{c}}\mapsto W_{\vec{a}}\otimes \left(W_{\vec{b}}\otimes
W_{\vec{c}}\right).$$ The picture below shows the two ways to get to
$I_{(\vec a,\vec b,\vec c)}\,W_{\vec a \plusd \vec b\plusd \vec c}$
where the ``trivial'' boundary Wilson line $I_{(a_\d,b_\d,c_\d)}$ has
charge $m_{\d}\left[a_\d+ b_\d +c_\d\over m_\d \right].$

 \vspace{1cm}
\setlength{\unitlength}{3cm}
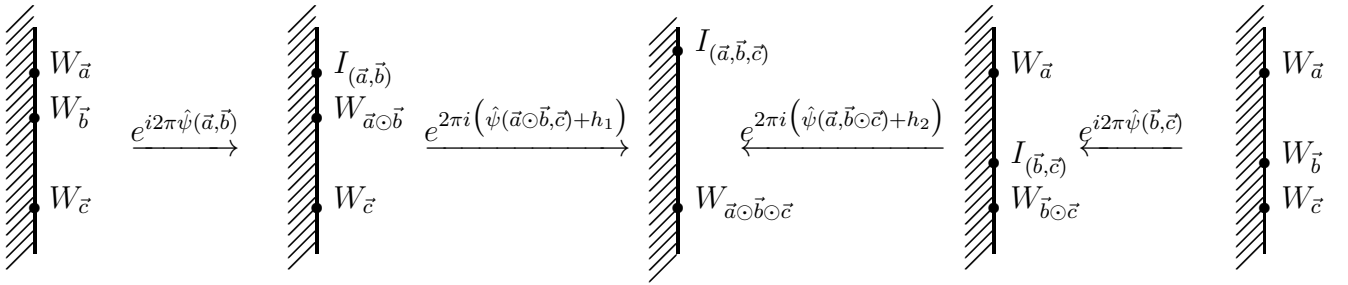
\begin{picture}(1,1)
 \linethickness{0.3mm}

\put(-0.125, 0){\line(0, 1){1}}
\multiput(-0.25,-0.07)(0,0.05){22}{\line(1,1){0.12}}

 \put(-0.125,
0.2){\circle*{0.05}}\put(-0.06, 0.2){$W_{\vec c}$}

%\put(3.325, 0.4){\circle*{0.05}}\put(3.375, 0.4){$W_{b_1}$}

\put(-0.125, 0.6){\circle*{0.05}}\put(-0.06, 0.6){$W_{\vec b}$}

\put(-0.125, 0.8){\circle*{0.05}} \put(-0.06, 0.8){$W_{\vec a}$}

\put(0.3, 0.5){$\underrightarrow{e^{i2\pi {\hat \psi}(\vec a,\vec
b)}}$}

\put(1.125, 0){\line(0, 1){1}}
\multiput(1,-0.07)(0,0.05){22}{\line(1,1){0.12}}

 \put(1.125,
0.2){\circle*{0.05}}\put(1.2, 0.2){$W_{\vec c}$}

%\put(1, 0.4){\circle*{0.05}}\put(1.1, 0.4){$W_{a_1}$}

\put(1.125, 0.6){\circle*{0.05}}\put(1.2, 0.6){$W_{\vec a \plusd
\vec b}$}

\put(1.125, 0.8){\circle*{0.05}} \put(1.2, 0.8){$I_{(\vec a,\vec
b)}$}

\put(1.6, 0.5){$\underrightarrow{e^{2\pi i\bigl({\hat \psi}(\vec a
\plusd \vec b, \vec c)+h_1\bigr)}}$}

\put(2.725, 0){\line(0, 1){1}} \multiput(2.6,
-0.125)(0,0.05){22}{\line(1,1){0.12}} \put(2.725,
0.2){\circle*{0.05}}\put(2.8, 0.2){$W_{\vec a \plusd \vec b \plusd
\vec c}$} \put(2.725, 0.9){\circle*{0.05}}\put(2.8, 0.9){$I_{(\vec
a,\vec b,\vec c)}$}

\put(3, 0.5){$\underleftarrow{e^{ 2\pi i\bigl({\hat \psi}(\vec a,
\vec b \plusd \vec c)+h_2\bigr)}}$}

 \put(4.125, 0){\line(0, 1){1}}
\multiput(4,-0.07)(0,0.05){22}{\line(1,1){0.12}}

 \put(4.125,
0.2){\circle*{0.05}}\put(4.2, 0.2){$W_{\vec b \plusd \vec c}$}

\put(4.125, 0.4){\circle*{0.05}}\put(4.2, 0.4){$I_{(\vec b,\vec
c)}$}

%\put(2.775, 0.6){\circle*{0.05}}\put(2.8, 0.6){$W_{\vec b}$}

\put(4.125, 0.8){\circle*{0.05}} \put(4.2, 0.8){$W_{\vec a}$}
\put(4.5, 0.5){$\underleftarrow{e^{i2\pi {\hat \psi}(\vec b,\vec
c)}}$}

\put(5.325, 0){\line(0, 1){1}}
\multiput(5.2,-0.07)(0,0.05){22}{\line(1,1){0.12}}

 \put(5.325,
0.2){\circle*{0.05}}\put(5.4, 0.2){$W_{\vec c}$}

%\put(3.325, 0.4){\circle*{0.05}}\put(3.375, 0.4){$W_{b_1}$}

\put(5.325, 0.4){\circle*{0.05}}\put(5.4, 0.4){$W_{\vec b}$}

\put(5.325, 0.8){\circle*{0.05}} \put(5.4, 0.8){$W_{\vec a}$}

 \end{picture}

 \vspace{1cm}

Here the phase $h_1$  arises from the rearrangement of ``trivial''
boundary Wilson lines from the configuration $I_{(\vec a,\vec b)}I_{(\vec a \plusd
\vec b, \vec c)}$ to $I_{(\vec a,\vec b,\vec c)}$ by moving
$I_{(a_\b \plusd b_\b, c_\b)}$ counterclockwise to the left of
$I_{(a_\c,b_\c)}$ for $\b < \c$:
$$h_1(\vec{a},\vec{b},\vec{c})=
 -\frac12 \sum_{\c>\b} m_\c \left[\frac{a_\c+b_\c}{m_\c}\right]m_\b
\left[\frac{(a_\b\plusd b_\b)+c_\b}{m_\b}\right] \sbh_{\b\c}.
$$
Meanwhile the phase $h_2$ arises by first moving $W_{\vec a}$
clockwise to the right of $I_{(\vec b,\vec c)}$ and then
rearranging trivial lines from $I_{(\vec b,\vec c)}I_{(\vec a, \vec
b\plusd \vec c)}$ to $I_{(\vec a,\vec b,\vec c)}$ by moving
$I_{(a_\b, b_\b \plusd c_\b)}$ counterclockwise to the left of
$I_{(b_\c,c_\c)}$ for $\b < \c$:
$$
h_2(\vec{a},\vec{b},\vec{c})=\frac12 \sum_{\c,\b} a_\c
m_\b\left[\frac{b_\b+c_\b}{m_\b}\right]\sbh_{\c\b}-\frac12\sum_{\c>\b} m_\c
\left[\frac{b_\c+c_\c}{m_\c}\right]m_\b \left[\frac{(b_\b\plusd
c_\b)+a_\b}{m_\b}\right] \sbh_{\b\c}.
$$
In this way we find that the boundary associator is a phase ${\mathbf
a}^{(bry)}_{a,b,c}=e^{2\pi i h_{bry}(a,b,c)}$ where
 \be \label{assoc_fin}h_{bry}(a,b,c)=h_2-h_1+\hat \psi(a,b\plusd
 c)-\hat \psi(a\plusd b,c)+\hat \psi(b,c)-\hat \psi(a,b).\ee

% It will prove to be convenient to add a co-boundary
%to $s_{sb}(A;b)$ and $h(a,b,c)$ (derived above) with
%$$k(a,b)=2\psi(a,b).$$

The last four terms in this formula are a coboundary, i.e. they can be removed by redefining the boundary tensor product of objects $W_{\vec{a}}$ and $W_{\vec{b}}$ by a phase $\hat\psi(a,b).$ Explicitly, this coboundary term is
 $$\half \sum_{\c >\b}a_\c \left[\frac{b_\b+
c_\b}{m_\b}\right]m_\b \sbh_{\b\c}+\half \sum_{\c
<\b}a_\c \left[\frac{b_\b+c_\b}{m_\b}\right]m_\b \sbh_{\b\c}-\sum_{\c<\b}c_\c \left[\frac{a_\b + b_\b}{m_\b}\right]m_\b
(\tv_\b,K^{-1}\tv_\c)$$
$$ +\frac12\sum_{\c>\b} m_\c
\left[\frac{a_\c+b_\c}{m_\c}\right]m_\b \left[\frac{(a_\b\plusd
b_\b)+c_\b}{m_\b}\right] \sbh_{\c\b}
-\frac12\sum_{\c>\b} m_\c
\left[\frac{b_\c+c_\c}{m_\c}\right]m_\b \left[\frac{(b_\b\plusd
c_\b)+a_\b}{m_\b}\right] \sbh_{\c\b}$$
Putting all these formulas together we get the boundary
associator:
\be \label{bryassoc_fin}h_{bry}(a,b,c)=\half \sum_{\d}
a_\d m_\d \left[{b_\d +c_\d\over m_\d}\right]\sb_{\d\d}-
 \sum_{\c<\b} c_\c m_\b
\left[{a_\b +b_\b\over m_\b}\right] \sb_{\c\b}\ee
$$+\sum_{\c \ne \b}a_\c m_\b\left[{b_\b +c_\b\over
m_\b}\right]\sb_{\c\b}.$$

A straightforward computation shows that the boundary associator (\ref{bryassoc_fin}) satisfies the pentagon identity. It is also independent of the projector $\Y$ which parameterizes the choice of the splitting.

The semi-braiding and the boundary associator are not unrelated: they satisfy the two boundary hexagon identities described in appendix B. One of them expresses the fact that the pair
$(W_{\pi(B)},s_{sb}(\cdot,B))$ is a well-defined object of the Drinfeld center of the category of boundary Wilson lines, and the other one says that the map which sends the object $W_B$ to the object of the Drinfeld center  $(W_{\pi(B)},s_{sb}(\cdot,B))$ respects the monoidal structure on the two categories. One can check that the semi-braiding and the boundary associator computed above satisfy the boundary hexagon identities.

It is important to realize that the boundary associator and the semi-braiding depend on the exact way the boundary tensor product is defined. If we redefine the product of $W_{\vec a}$ and $W_{\vec b}$ by a phase $k(a,b)$,  the boundary associator and the semi-braiding change as follows:
\begin{align}
h'_{bry}(a,b,c)&=h_{bry}(a,b,c)+k(a, b\plusd c)-k(a\plusd b,c)+k(b,c)-k(a,b),\\
s'_{sb}(a,B)&=s_{sb}(a,B)+k(\pi(B),a)-k(a,\pi(B)).
\end{align}
We will say that $(h'_{bry},s'_{sb})$ differs from $(h_{bry},s_{sb})$ by a coboundary $k(a,b)$.
The new boundary associator still satisfies the pentagon identity, and the new boundary associator and the semi-braiding together still satisfy the boundary hexagon identities. Such a redefinition of the boundary tensor product replaces the monoidal category of boundary Wilson lines with an equivalent one and should be regarded as physically trivial.
Note that the semi-braiding (\ref{semi_fin}) and the boundary associator (\ref{assoc_fin}) differ from
$h^{(0)}_{bry}=h_2-h_1$ and $s^{(0)}_{sb}$ by a coboundary  $\hat\psi(\vec a, \vec b).$ This  coboundary makes $h_{bry}$ and $s_{sb}$ independent of the projector $\Y.$ This independence is less obvious if we use $h^{(0)}_{bry}$ and $s^{(0)}_{sb}$ because under a change of $\Y$ they change by a coboundary.

It is useful to note that the boundary associator can be further simplified by adding a
coboundary corresponding to
$$k(a,b)=\sum_{\c > \b}a_{\c}b_{\b} (\tv_{\c},K^{-1}\tv_{\b}).$$
The new boundary associator is
\be \label{assoc_new}h'_{bry}(\vec a,\vec b,\vec c)=\half \sum_{\d}
a_\d m_\d \left[{b_\d +c_\d\over m_\d}\right]
(\tv_{\d},K^{-1}\tv_{\d})+\sum_{\c < \b}a_\c m_\b\left[{b_\b
+c_\b\over m_\b}\right](\tv_{\c},K^{-1}\tv_{\b}).\ee
The new semi-braiding is
\be \label{semi_new} s'_{sb}(\vec a;\vec B)=-\half \sum_{\c=1}^n b_\c
a_\c(\tv_\c, K^{-1}\tv_\c) -\sum_{\c >\a}b_\a a_\c (\tv_\a,
K^{-1}\tv_\c)\ee
$$-\sum_{\hat j=n+1}^N\sum_{\c=1}^n B_{\hat j} a_\c
(r_{\hat j}, P^t\tv_\c)- \sum_{\a,\c} a_{\c}
m_{\a}\tB_{\a}(\tv_\c,K^{-1}\tv_\a).$$

Finally let us discuss how the boundary associator depends on various arbitrary choices we have made. The formulas obviously depend on the ordering of the generators $v_1,\ldots,v_n$. As in the bulk case, it is easy to check that changing the order modifies the boundary associator and the semi-braiding by a coboundary.  We also chose a lift of the generators of $\D_0$ to $\D$. We claim that changing a lift also adds a coboundary. Indeed, changing a lift amounts to a replacement
$$\tv_\c\mapsto \tv_\c
+KPf_\c,\quad f_\c \in \Lambda_0\otimes\QQ,\quad KPf_\c \in \Lambda^*.$$
Simultaneously we need to perform a shift
$$\sum_{\hat j}B_{\hat j}r_{\hat j} \mapsto
\sum_{\hat j}B_{\hat j}r_{\hat j}-\sum_{\c}b_\c f_{\c}$$ so that
$\sum_{j=1}^N B_j e_j$ is unchanged. Then a short computation shows that this transformation adds to $(h'_{bry},s'_{sb})$ a coboundary corresponding to
$$
k(a,b)=-\sum_{\alpha<\gamma}a_\alpha b_\gamma (\tv_\alpha,P f_\gamma).
$$
Alternatively, it is easy to check that $(h_{bry}^{(0)},s_{sb}^{(0)})$, which differs from $(h'_{bry}, s'_{sb})$ by a coboundary, is unchanged when we do the above transformation.
To summarize, the boundary associator and the semi-braiding do not depend on arbitrary choices, up to a coboundary.

Despite using the ``cohomological'' terminology, we have not explained what cohomology theory is relevant here. It is easy to supply such a theory if we ignore the semi-braiding. Then the boundary associator phase defines a 3-cochain in the standard complex which computes the cohomology of the group $\D_0=\D/\sM$ with values in the trivial module $\QQ/\ZZ$. The pentagon identity says that this 3-cochain is a 3-cocycle, and identifying boundary associators differing by a coboundary is exactly the same as identifying cohomologous 3-cocycles. Our computations in this section show how to associate a canonical element if $H^3(\D/\sM,\QQ/\ZZ)$ to any Lagrangian subgroup $\sM$ of a finite abelian group $\D$ equipped with a quadratic function $\sq$. In fact, it is easy to see that in all our computations we never used the fact that $\sM$ is coisotropic, so the formulas are valid for an arbitrary isotropic subgroup of $(\D,\sq)$. We discuss the relevance of the coisotropic condition in section 8.

\subsection{Examples}
Let us illustrate the above discussion by two simple examples. The first one involves the topological
boundary condition introduced in section 4.4. The bulk gauge group is
$U(1)^{2m}$ and the block-diagonal Chern-Simons coupling matrix is given by
(\ref{CSexample1}). The bulk discriminant group $\D$ consists of pairs $({\bf x},{\bf y})$ where ${\bf x}$ and ${\bf y}$ are elements of $\ZZ^m$ defined up to addition of elements of the form $\KK\lambda$, where $\lambda$ is an arbitrary element of $\ZZ^m$. Elements of $\D$ label bulk Wilson lines. The boundary gauge group is the diagonal
$U(1)^m$ subgroup. The charge of a Wilson line ending on the boundary must be in the kernel of $P^t$, therefore it must have the form $({\bf x},-{\bf x}),$ where ${\bf x}\in\ZZ^m$ and is again defined up to addition of $\KK\lambda$. Elements of this form define the subgroup  $\sM$ in $\D$. The boundary discriminant group $\D_0=\D/\sM$ is isomorphic to $\sM$, and in fact we have $\D\simeq \D_0\oplus \D_0$.

We  can always write $\D_0$ as $\oplus_{\c=1}^n \ZZ_{m_\c}$.  Let $v_\c$, $\c=1,\ldots, n$, be an ordered set of generators of $\D_0$. We can lift them to generators of $\D$ by writing
$$
e_\c=(v_\c,0),\quad \c=1,\ldots,n.
$$
Generators of $\sM$ can be taken to have the form
$$
e_{n+\c}=(v_\c,-v_\c).
$$
Together they generate the whole $\D$. Let $\tv_\c$ be some lift of $v_\c$ to $\ZZ^m$. That is, we have $m_\c \tv_\c=\KK f_\c$ for some $f_\c\in\ZZ^m$. Then we have
$$
\te_\c=(\tv_\c,0),\quad \te_{n+\c}=(\tv_\c,-\tv_\c),\quad \c=1,\ldots,n.
$$

Boundary Wilson lines are labeled by vectors ${\vec a}=(a_1,\ldots,a_n)$ where $a_\c$ is an integer in the interval $[0,m_\c-1]$. The boundary associator in this example is given by
\be \label{assoc_ex1}h'_{bry}(\vec a,\vec b,\vec c)=\half \sum_{\d}
a_\d m_\d \left[{b_\d +c_\d\over m_\d}\right]
(\tv_{\d},\KK^{-1}\tv_{\d}).\ee
Note that this is the same as the bulk associator in the $U(1)^m$ Chern-Simons theory with the symmetric blinear form $\KK$. This happens because the boundary condition we are considering can be reinterpreted (via the folding trick, see section 7) as a trivial surface operator in the $U(1)^m$ Chern-Simons theory. Boundary Wilson lines are then reinterpreted as bulk Wilson lines in this theory.

Bulk Wilson lines are labeled by vectors $\vec{B}=(B_1,\ldots,B_{2n})$, where $\forall \c\in\{1,\ldots,n\}$ $B_\c$ and $B_{n+\c}$ are integers in the interval $[0,m_\c-1]$.
The semi-braiding is
\be \label{semi-ex} s'_{sb}(\vec
a;\vec B)=-{1\over 2}\sum_{\c=1}^n a_\c B_\c (\tv_\c,\KK^{-1} \tv_\c)-\sum_{\c,\b}
B_{n+\b}a_\c (\tv_\b,\KK^{-1}\tv_\c)-\sum_{\c >\a}a_\c B_\a (\tv_\a,\KK^{-1}\tv_\c).\ee

Note that in this example the boundary associator is a 2-torsion, i.e it is $\pm 1$ in the multiplicative notation. In general the boundary associator is not a 2-torsion. To show this, consider a $U(1)\times U(1)$ theory with
$$
K=\begin{pmatrix} 4 & -5 \\ -5 & 4\end{pmatrix}.
$$
The bulk discriminant group in this example is $\D\simeq \ZZ_9$, and the quadratic function $\sq:\D\ra\QQ/\ZZ$ is given by
$$
\sq(x)=-\frac{2}{9}x^2,\quad x\in \ZZ/9\ZZ.
$$
We can take the generator of $\D$ to be $\te=(1,0)$, since
$$
K\begin{pmatrix} -4\\ -5 \end{pmatrix}=9e.
$$
Consider the boundary condition defined by
$$
P=\begin{pmatrix} 2 \\ 1\end{pmatrix}.
$$
The subgroup $\sM\subset\D$ is generated by the vector $(1,-2)$ and is isomorphic to $\ZZ_3$ since
$$
\begin{pmatrix} 3 \\ -6\end{pmatrix}=K\begin{pmatrix} 2 \\ 1 \end{pmatrix}.
$$
The boundary discriminant group $\D_0=\D/\sM$ is isomorphic to $\ZZ_3$ as well and is generated by $\tv=(1,0)$. Boundary Wilson lines are labeled by $a\in\{0,1,2\}$. The boundary associator is
$$
h'_{bry}(a,b,c)=-\frac{1}{3} a\left[\frac{b+c}{3}\right].
$$
It takes values in third roots of unity.\footnote{Incidentally, this implies that the monoidal category of boundary line operators in this case does not admit any braided monoidal structure.}

Bulk Wilson lines are labeled by an integer $B\in\{0,1,\ldots,8\}$. The semi-braiding is
$$
s'_{sb}(a,B)=\frac{2}{9} aB+\frac{2}{3}a\left[\frac{B}{3}\right].
$$

\subsection{Reduction on an interval}\label{sec:reduction}

So far we have been studying boundary line operators on a particular boundary. We may also study line operators separating two different boundary conditions; we may call them boundary-changing line operators. From the mathematical viewpoint boundary conditions in a 3d TFT are objects of a 2-category. Boundary line operators separating boundary conditions $\sA$ and $\sB$ are 1-morphisms from $\sA$ to $\sB$, while local operators sitting at the junctions of two such line operators are 2-morphisms. So far we have been studying 1-morphisms from an object to itself, and now we turn to more general 1-morphisms.

In the case of abelian Chern-Simons theory boundary conditions are labeled by Lagrangian subgroups of the discriminant group $\D$.
To describe boundary-changing line operators between Lagrangian subgroups $\sM_1$ and $\sM_2$ it is useful to compactify our 3d TFT on an interval with boundary conditions corresponding $\sM_1$ and $\sM_2$ on the two ends. Boundary-changing line operators are in one-to-one correspondence with boundary conditions in the resulting 2d TFT. More precisely, boundary-changing line operators from $\sM_1$ to $\sM_2$ form a category $\Hom(\sM_1,\sM_2)$, and this category is isomorphic to the category of D-branes in the effective 2d TFT \cite{KRS,Kap:ICM}.

Consider abelian Chern-Simons theory on $M_3=\Sigma\times [0,L]$ with boundary conditions $(P_1,R_1)$ at $x^3=0$
and $(P_2,R_2)$ at $x^3=L.$ Here $P_i$ determines the boundary gauge group $\TT_{\Im P_i}$, and $R_i$ is the splitting of the corresponding exact sequence (\ref{splittingvect}). The splittings $R_i$ enter the boundary conditions for $A_{\perp}=A_3$, the Lagrange multiplier field $B$ and the Faddeev-Popov ghosts. It is actually easier to perform the reduction without fixing the gauge. Since at $x^3=0$ the gauge group is reduced to $\TT_{\Im P_1}$ and at $x^3=L$ it is reduced to $\TT_{\Im P_2}$, the 2d gauge group will be the isotropic subgroup
$$
\sA_{12}=\TT_{\Im P_1}\cap\TT_{\Im P_2}.
$$
$\sA_{12}$ is not necessarily connected, i.e. it may be a product of a torus and a finite group. Its Lie algebra is an isotropic subspace of $\g_\Lambda$:
$$
\g_{12}=(\Im P_1\otimes\RR)\cap (\Im P_2\otimes\RR).
$$
Another way to explain this is to say that only the constant mode of $A_{\para}$ survives the reduction, and since at $x^3=0$ and $x^3=L$ the field $A_{\para}$ must lie in $\Im P_1\otimes\RR$ and $\Im P_2\otimes\RR$, respectively, the constant zero mode must lie in the intersection of these two vector spaces.

The component $A_3$ gives rise to a 2d scalar
$$
\varphi=\int_0^L dx^3 A_3.
$$
In terms of this scalar the 2d effective action becomes
$$
S_{12}=\frac{i}{2\pi} \int K(\varphi,dA_{\para}).
$$
The scalar field $\varphi$ is not gauge-invariant: under an $x^3$-dependent gauge transformation $\alpha$ it shifts by $\alpha(L)-\alpha(0)$. Since $\alpha(0)\, {\rm mod}\, 2\pi\Lambda$ and $\alpha(L)\, {\rm mod}\, 2\pi\Lambda$ belong to $(\Im P_1)\otimes\RR$ and $(\Im P_2)\otimes\RR$, respectively, this means that $\varphi\in\g_\Lambda$
 is defined modulo arbitrary elements of the subgroup
 $$(\Im P_1\otimes \RR) \oplus (\Im P_2\otimes \RR) \oplus 2\pi \Lambda.$$
Equivalently, we may regard $\varphi$ as living in the quotient torus
$$
\TT_\Lambda/\TT_{\Im P_1\oplus \Im P_2}.
$$
The action $S_{12}$ is non-degenerate because $K$ is a non-degenerate pairing between the vector spaces
$$
\g_\Lambda/((\Im P_1\otimes\RR) \oplus (\Im P_2\otimes\RR))
$$
and
$$
(\Im P_1)\otimes\RR\cap (\Im P_2)\otimes\RR.
$$
This follows from the fact that $\Im P_1$ and $\Im P_2$ are Lagrangian subspaces of $\g_\Lambda$.

The effective 2d TFT is thus a 2d version of the BF theory, with the periodic scalar field $\varphi$ playing the role of $B$. The construction of branes in this theory is fairly obvious: one places on the boundary a quantum-mechanical degree of freedom taking values in some representation of the gauge group $\sA_{12}$ and couples it to the restriction of the gauge field. Thus branes are labeled by elements of the abelian group $\Hom(\sA_{12},U(1))$. In the special case $P_1=P_2=P$ the group $\sA_{12}$ is simply the torus $\TT_{\Lambda_0}$ where $\Lambda_0=\Im P$, so we recover the result that boundary line operators are (direct sums of) Wilson lines for the boundary gauge group $\TT_{\Lambda_0}$.

Note that since $U(1)$ is a divisible group, any element of $\Hom(\sA_{12},U(1))$ is a restriction of an element of $\Hom(\TT_{\Lambda},U(1))=\Lambda^*$. That is, the group of charges of boundary-changing line operators is a quotient of the group of bulk charges. The kernel of the quotient map consists of bulk charges which vanish on $\sA_{12}$. Equivalently, one might say that the kernel consists of bulk Wilson lines which are neutral with respect to $\sA_{12}$ and therefore may end on the junction between the two boundaries.

We should also take into account isomorphisms between boundary-changing line operators arising from monopole operators. That is, if wish to identify isomorphic boundary-changing line operators, we should further quotient by the subgroup $\Im K$. If we reverse the order of quotients, we get the following description of the set of isomorphism classes of boundary-changing line operators: it is the quotient of the bulk discriminant group $\D$ by the subgroup generated by $\sM_1$ and $\sM_2$. This subgroup describes bulk Wilson lines which may end at the junction of boundary conditions defined by subgroups $\sM_1$ and $\sM_2$.

A boundary-changing line operator from $\sM_1$ to $\sM_2$ can be fused ``from the left'' with a boundary line operator on the boundary $\sM_1$ or ``from the right'' with a boundary line operator on the boundary $\sM_2$. This makes the category of boundary-changing line operators a bi-module category over the monoidal categories of boundary line operators $\Hom(\sM_1,\sM_1)$ and $\Hom(\sM_2,\sM_2)$. It is easy to see how objects map under this action: fusing a boundary line operator labeled by $x\in \D/\sM_i$, $i=1,2$ with a boundary-changing line operator $z\in \D/(\sM_1+\sM_2)$ gives a boundary-changing line operator $x\circ z=x+z\in \D/(\sM_1+\sM_2)$. A complete description of the bi-module category structure requires describing an an ``associator'' between $(x\circ y)\circ z$ and $x\circ(y\circ z),$
where $x,y\in \D/\sM_i$ and $z\in  \D/(\sM_1+\sM_2)$. We leave this for future work.

\section{More general boundary conditions}\label{sec:generalbdries}

\subsection{Boundary conditions with a disconnected boundary gauge group}

So far we have been assuming that the boundary gauge group is connected and therefore is a torus. More generally, the boundary gauge group may have several components each of which is a torus. In this section we study the corresponding boundary conditions and line operators on such boundaries.

In this section it is convenient, instead of imposing boundary conditions on the gauge field $A$ ``by hand'', to introduce Lagrange multiplier fields living on the boundary whose equations of motion enforce the desired boundary conditions. First let us see how this works in the case of a connected boundary  gauge group. Let $\Lambda_0$ be a Lagrangian sublattice in $\Lambda$. We can try to enforce the condition that on the boundary the holonomy of $A$ lies in $\TT_{\Lambda_0}$ by introducing a Lagrange multiplier field $\B$ living on the boundary and adding a boundary action
$$
\frac{i}{2\pi}\int_{\partial M} V(\B, dA).
$$
The field $\B$ takes values in a torus $\TT_\M$ whose dimension is half the rank of $\Lambda$, and $\Phi$ is a fintely-generated free abelian group whose rank is half the rank of $\Lambda$. To write down the action we need to lift $\B$ to a field valued in $\M\otimes \RR$. In order for the action to be well-defined the coupling matrix $V$ must be integral, i.e. it should be an element of the abelian group $\Hom(\M,\Lambda^*)$.
We will denote by $V^t$ the dual homomorphism from $\Lambda$ to $\M^*$.

The equation of motion for $\B$ gives a constraint
$$
V^t(dA)=0.
$$
Since we want the constraint to enforce the correct boundary condition for $dA$, we need to require, as a minimum,
$$
\ker V^t=\Lambda_0.
$$
To ensure that the gauge group on the boundary is $\TT_{\Lambda_0}$ we need to impose a stronger constraint that the boundary holonomies of $A$ belong to the torus $\TT_{\Lambda_0}$. The condition for this is as follows. Since $\ker V^t=\Lambda_0$, $\Im\, V$ annihilates $\Lambda_0$, i.e. $\Im\, V\subset \ker P^t$, where $P^t$ is the projection from $\Lambda^*$ to $\Lambda_0^*$. Boundary holonomies of $A$ belong to $\TT_{\Lambda_0}$ if and only if $\Im\, V=\ker P^t$. Indeed, integration over $\B$ enforces a constraint that contour integrals of $A/2\pi$ take integral values on $\Im\, V$. On the other hand, we would like to require that contour integrals of $A/2\pi$ take integral values on the sublattice $(\Lambda/\Lambda_0)^*$ of $\Lambda^*$, i.e. on $\ker P^t$. Clearly, for the two conditions to be equivalent, we must have $\Im\, V=\ker P^t$.

If $\Im\, V$ is a proper sublattice of $\ker P^t$, then the quotient $\ker P^t/\Im V$ is a finite abelian group of order $d>1$. In such a case the boundary holonomy of $A$ does not necessarily lie in the torus $\TT_{\Lambda_0}$, but the $d^{\rm th}$ power of the holonomy does.  That is, the boundary gauge group in this more general case is disconnected and consists of several Lagrangian tori inside the torus $\TT_\Lambda$. But this still ensures that $\delta A$ takes values in a subspace isotropic with respect to $K$. Thus we can drop the constraint $\ker P^t=\Im V$, thereby allowing disconnected boundary gauge groups.

Let us describe the boundary gauge group in more detail. It is a subgroup of the bulk gauge group $\TT_\Lambda$. Any such subgroup is a kernel of some homomorphism from
$\TT_\Lambda$ to some other torus $\TT_\Gamma$ induced by a surjective homomorphism $h$ from $\Lambda$ to $\Gamma$. In other words, if we regard $h$ as a linear map from $\Lambda^\RR$ to $\Gamma^\RR$, then an element $x\in\Lambda_\RR/(2\pi\Lambda)$
belongs to the subgroup iff $hx/2\pi$ belongs to $\Gamma$. In our case, the constraint on the holonomy $x$ of $A/2\pi$ says that $V^t x$ belongs to the lattice $\M^*$. This means that $\Gamma=\M^*$ and $h=V^t$. That is, the boundary gauge group is the kernel of the Lie group homomorphism $V^t_\TT:\TT_\Lambda\ra\TT_{\M^*}$ induced by the homomorphism $V^t:\Lambda\ra \M^*$.

So far the only constraint on $V$ was that $\ker V^t$ is a Lagrangian sublattice of $\Lambda$. There is one more constraint on $V$ which arises from gauge invariance. Gauge transformations in the bulk take values in the torus $\TT_\Lambda$, but on the boundary they are constrained to lie in the Lie subgroup $\ker V^t_\TT$, as explained above. Under such a gauge transformations the bulk action varies by a boundary term
$$
-\frac{i}{2\pi}\int_{\partial M} K(f,dA).
$$
The gauge variation of the boundary term is
$$
\frac{i}{2\pi}\int_{\partial M} V(\delta \B, dA)
$$
Note that while integration over $\B$ produces a delta-functional which constrains $dA$ to lie in $\Lambda_0\otimes\RR$, before we integrated over $\B$ the form $dA$ can take values in $\Lambda$.

Let us consider $f$ which lies in the identity component $\TT_{\Lambda_0}$ of the boundary gauge group. In order to cancel the variation of the bulk term we need to let $B$ transform as well. Comparing the two variations we see that we must have
$$
\delta \B=W_\TT(f),
$$
where the Lie group homomorphism $W_\TT:\TT_{\Lambda_0}\ra\TT_\M$ is induced by a homomorphism  $W:\Lambda_0\ra\M$. For the total gauge variation to vanish we must have
$$
K_0=VW.
$$
Here $K_0:\Lambda_0\ra\Lambda^*$ is a restriction of $K$ to $\Lambda_0$.

We can write this constraint in a more convenient form by noting that for any $x\in\Lambda_0$ $K_0(x)$ annihilates $\Lambda_0$. Similarly, for any $y\in \M$ $V(y)$ annihilates $\Lambda_0$. Hence we can interpret $K_0$ as a homomorphism from $\Lambda_0$ to $(\Lambda/\Lambda_0)^*$ , and we can interpret $V$ as a homomorphism from $\M$ to $(\Lambda/\Lambda_0)^*$. Then the equation $K_0=VW$ becomes an equation for three integral square matrices.

For a fixed $K$ and $\Lambda_0$ we may look for possible $\M,$ $V$ and $W$ by trying to factorize $K_0$ as a product of two integral square matrices $V$ and $W$. A slightly different way to phrase this is as follows. $K_0$ gives an embedding of $\Lambda_0$ into $(\Lambda/\Lambda_0)^*$ (as a sublattice of maximal rank).  Finding $V$ and $W$ such that $K_0=VW$ is equivalent to finding  a sublattice $\M$ of  $(\Lambda/\Lambda_0)^*$ which contains $K_0(\Lambda_0)$. There are two obvious solutions. We can take $\M=(\Lambda/\Lambda_0)^*$, $V=1$ and $W=K_0$. Or we can take $\M=\Lambda_0,$ $V=K_0$ and $W=1$. We will call them the minimal and the maximal solutions. The boundary gauge group for the minimal solution is the torus $\TT_{\Lambda_0}$. For the maximal solution the boundary gauge group is the kernel of the Lie group homomorphism $\TT_\Lambda\ra \TT_{\Lambda_0^*}$ induced by the homomorphism $\pi\circ K:\Lambda\ra\Lambda_0^*$. It is disconnected in general, with the identity component being $\TT_{\Lambda_0}$. The minimal and maximal solutions coincide if and only if $K_0=1$.

%\subsection{Some examples}

\subsection{Wilson lines}
Now let us analyze which bulk Wilson lines may terminate on the
boundary. Let us start with a simple case of an abelian gauge theory
with gauge group $G=U(1)\times U(1)$ and Chern-Simons levels $2N$
and $-2N$. That is, we take the matrix $K$ to be
$$
K=\begin{pmatrix} 2N & 0 \\ 0 & -2N\end{pmatrix}.
$$
We can identify both $\Lambda$ and $\Lambda^*$ with $\ZZ\oplus \ZZ$.
We choose $\Lambda_0$ to consist of elements of $\ZZ\oplus\ZZ$ of
the form $(n,n)$, so that the torus $\TT_{\Lambda_0}$ is the
diagonal subgroup of $U(1)\times U(1)$. The quotient
$(\Lambda/\Lambda_0)^*$ is one-dimensional and can be identified
with the sublattice of $\ZZ\oplus\ZZ$ of the form $(n,-n)$. The
image of $K_0$ can be identified with the sublattice consisting of
the elements of the form $(2N,-2N)$. The matrices $V$ and $W$ are
one-by-one in this case and satisfy $VW=2N$. Thus choices of $\M$
are in 1-1 correspondence with divisors of $2N$. If $v$ is such a
divisor, then we set $V=v$ and $W=2N/v=w$. The minimal solution is
$v=1, w=2N$, the maximal solution is $v=2N, w=1$.

The boundary gauge group is a subgroup of $U(1)\times U(1)$ defined
by the condition $v(x-y)=0$,
 where $x$ and $y$ take values in $\RR/2\pi\ZZ$.  The solution of this constraint  is
$$
y=x+\frac{2\pi \ell}{ v},\quad \ell\in \ZZ/v\ZZ
$$
This shows that the boundary gauge group is isomorphic to
$U(1)\times \ZZ_v$.

The identity component of the gauge group is $U(1)\simeq
\RR/2\pi\ZZ$. Under the corresponding gauge transformations the
boundary scalar $\B$ transforms as
$$
\B\mapsto \B+w f,\quad f\in C^\infty(\partial M,\RR/2\pi\ZZ).
$$
We still have not determined the transformation law of $\B$ under
the discrete part of the boundary gauge group. Consider a bulk gauge
transformation whose value on the boundary is given by
$$
(f_1,f_2)=(0,2\pi \ell/v)
$$
The variation of the bulk action under such a transformation is
$$
\frac{2N i\ell}{v}\int_{\partial M} dA_2=iw \ell \int_{\partial M} dA_2.
$$
Since $\int dA_2$ is quantized in units of $2\pi$, the exponential
of this term is $1$. Now let us look at the boundary action. The
most general transformation law for $\B$ under the discrete gauge
transformations is
$$
\B\mapsto \B+2\pi m\ell /v,
$$
where $\ell\in\ZZ/v\ZZ$ is the parameter of gauge transformation and
$m$ determines the choice of the transformation law for $\B$. It is
easy to see that the variation of the boundary action is also an
integral multiple of $2\pi i$ and so no constraint on $m$ arises
from gauge-invariance. We will see below that $m$ is
constrained by locality considerations. This will lead to an
additional constraint on the choice of $v$ and $w$.

 A Wilson line
can terminate on the boundary if and only if the charge of the
end-point of the Wilson line can be screened by a local operator
sitting on the boundary. Recall also that there are bulk monopole
operators which carry bulk electric charges of the form $(2N \ZZ,2N
\ZZ)$,  so we may regard bulk electric charges as defined modulo
$2N$. On the boundary we have additional local operators built from
$\B$. There are both order and disorder operators of this kind. The
order operators have the form
$$
e^{i\nu \B},\quad \nu\in\ZZ.
$$
We will call the integer $\nu$ the momentum. The charge of such an
operator with respect to the identity component of the boundary
gauge group is $\nu w$. A disorder operator is defined by the
condition that as one goes around the insertion point $\B$ winds
$\mu$ times:
$$
\B\mapsto\B+2\pi\mu,\quad \mu\in\ZZ.
$$
We will call $\mu$ the winding number. In the presence of a disorder
operator $d\B$ is a closed 1-form with a singularity at the
insertion point such that $d(d\B)=2\pi\mu \delta^2(x-x_0)$. The
variation of the boundary action with respect to the identity
component of the boundary gauge group vanishes. Hence the charge of
the end-point of a Wilson line has to be screened by the order
operator alone. This means that such a Wilson line must have a
charge of the form $X=(X_1,X_2)$ with $X_1+X_2=0\, {\rm mod}\, w$.
For the minimal solution $v=1,w=2N$ this constraint reduces to
$X_1+X_2=0\, {\rm mod}\, 2N$, i.e. trivial charge with respect to
the boundary gauge group, modulo screening by bulk monopoles. This
agrees with what we obtained in section 4.1 without introducing the
boundary scalar $\B$.

For $v\neq 1$ the boundary gauge group also has a factor
$\ZZ_v$, and one needs to require that the $\ZZ_v$ charges of the
endpoint of the Wilson line and the order operator cancel. If the
$\ZZ_v$ charge of $\B$ is $m\, {\rm mod}\, v$, then this condition
is
$$
X_2=\nu m+\mu v
$$
where $\nu\in\ZZ$ is defined by the relation
$$
X_1+X_2=\nu w,
$$
and $\mu$ is an arbitrary integer.

So far $m$ was left undetermined. We will now argue that
diffeomorphism invariance of the boundary condition on the quantum
level requires $2m=w \mod 2v$. Consider $M_3$ of the form $S^2\times
I$. The boundary of this manifold has two connected components, so
let us consider a Wilson line which begins at one boundary
components and ends on the other one. On the quantum level Wilson
lines have to be regularized by thickening them into ribbons; this
is called a choice of framing. Topological correlators for bulk
Wilson loops depend on this choice: twisting a ribbon of charge $X$
by a full turn multiplies the correlator by \cite{Witten}
\begin{equation}\label{twistribbon}
\exp(i\pi K^{-1}(X,X))=\exp(2\pi i \sq(X)).
\end{equation}
But in the situation described above such a twist can be
 undone by a rotation of the boundary $S^2$. Hence we must require that the
  correlator be independent of the choice of framing. In the simple example we are considering this gives the following condition:
$$
X_1^2-X_2^2=0 \mod 4N.
$$
Expressing $X_1$ and $X_2$ in terms of $\nu$ and $\mu$ we get
$$
\nu^2 w (w-2m)-2\mu\nu vw=0\mod 4N.
$$
Since $vw=2N$, this implies
$$
w-2m=0 \mod 2v.
$$
Therefore $w$ must be even and $m=w/2 \mod v$. Since $w$ is
even and $v=2N/w$, we see that $v$ must actually be a divisor of
$N$, not just a divisor of $2N$.

With this choice of $m$ the constraints on $X_1,X_2$ may be rewritten in a more symmetric way:
$$
X_1+X_2=0 \mod w,\quad X_1-X_2=0 \mod 2v.
$$
These constraints ensure that the endpoints of Wilson lines are mutually local.
 Indeed, the Aharonov-Bohm phase arising from transporting a charge $X$ around a charge $Y$ is
\begin{equation}\label{ABphase}
\exp(-2\pi i \sb(X,Y))=\exp(-2\pi i (\sq(X+Y)-\sq(X)-\sq(Y))).
\end{equation}
Since our constraints ensure that the phase (\ref{twistribbon}) is
trivial for all Wilson lines ending on the boundary, the phase
(\ref{ABphase}) is also trivial.

Let us discuss another example of a generalized boundary condition in a $U(1)\times U(1)$ Chern-Simons theory, with a nondiagonal form $K$:
$$
K=\left(\begin{tabular}{cc}$4$ & $-4$\\
$-4$ & $0$\end{tabular}\right).
$$
Consider a boundary condition defined by
$$
P=\left(\begin{tabular}{c}$2$ \\
$1$\end{tabular}\right)
$$
We can solve the condition $K_0=VW$ by letting
$$V^t=v(1, -2),\quad W={4\over v},$$
where $v$ divides $4$.
On the boundary a gauge transformation $(f_1,f_2)$ must
satisfy $v(f_1-2f_2) \in 2\pi \ZZ$. The general solution has the form
$$f_1=2f_2+{2\pi \alpha \over v}\quad \alpha=0,\ldots, v-1.$$ That is,  the boundary gauge group is isomorphic to $U(1)\times \ZZ_v.$ Let the charge
of a Wilson line ending on the boundary be $X=(X_1, X_2),$ then
the screening of the $U(1)$ charge on the boundary requires
\be\label{new}
2X_1+X_2=W\nu,\quad \nu\in\ZZ.
\ee
The locality condition in
this case reads \be \label{loc} X_2^2+2X_1X_2=0 \ \text{mod}\, 8.\ee
The screening of the $\ZZ_v$ charge on the boundary requires
 \be\label{screen_iv} X_1=m\nu +\mu v\ee for some integer $\mu$. Now expressing $X_1,X_2$ in terms of $\mu,\nu$ we get
from (\ref{loc}):
$$ W(W-2m)=0\  \text{mod}\, 8.$$ So we conclude
$${4\over v}-2m=0 \  \text{mod}\, 2v$$
This equation determines the allowed values of $v$ and the corresponding discrete charge $m$ of the field $\varphi$:
\begin{itemize}
\item
$v=2,\quad m=1\ \text{mod}\, 2$

\item $v=1,\quad m=0$
\end{itemize}

In the case $v=2,m=1$ the charge $X=(X_1,X_2)\in\Lambda^*$ of a Wilson line ending on the boundary must satisfy
 $$
 X_2=0\ \text{mod}\, 4.
 $$
After taking into account  identifications due to monopoles, we may regard both $X_1$ and $X_2$ as integers modulo $4$.
Then the condition on $(X_1,X_2)$ becomes simply $X_2=0$.

On the other hand, in the case $v=1,m=2$ the charge of the Wilson line ending on the boundary must satisfy
 $$X_2+2X_1=0.$$

 Note that the bulk discriminant group in this case is $\D=\ZZ_4\times\ZZ_4$, with the quadratic function
 $$
 \sq(X)=-\frac{1}{8} (X_2^2+2X_1X_2)
 $$
 Both in the case $v=1$ and in the case $v=2$ charges of Wilson lines ending on the boundary generate a subgroup $\sM$ of $\D$ isomorphic to $\ZZ_4$. It is easy to check that in both cases $\sM$ is Lagrangian with respect to $\sq$.\footnote{Isotropicity of $\sM$ is equivalent to the locality constraint on the charges of Wilson lines ending on the boundary. The fact that $\sM$ is coisotropic is less trivial, and we do not know of a general proof.} This lends support to the proposal that an arbitrary Lagrangian subgroup of $(\D,\sq)$ defines a valid topological boundary condition.

Let us indicate how to extend this analysis
to a general abelian Chern-Simons theory. An order operator on the
boundary has the form
$$
e^{i\nu(\B)},
$$
where $\nu$ is an element of the lattice
$\M^*=\Hom(\M,\ZZ)=\Hom(\TT_\M,U(1))$. The charge of such an
operator with respect to the identity component  of the boundary
gauge group $\TT_{\Lambda_0}$ is $W^t(\nu)\in\Lambda_0^*$. Thus the
charge $X$ of a Wilson line endpoint can be screened by such an
order operator only if $P^t X$ lies in the lattice $\Im\, W^t$. In
general $\Im\, W^t$ is a sublattice of maximal rank in
$\Lambda_0^*$, so this is a nontrivial condition.

We also need to require that the charge of the Wilson line with
respect to the discrete part of the boundary gauge group be screened
by the order operator. The transformation law of $\B$ under the
boundary gauge group $\ker V_\TT^t$ is specified by a homomorphism
$\tilde W:\ker V_\TT^t\ra \TT_\M$. The restriction of $\tilde W$ to
the torus $\TT_{\Lambda_0}$ must coincide with the homomorphism
$W_\TT:\TT_{\Lambda_0}\ra \TT_\M$ induced by $W:\Lambda_0\ra \M$.
Given a choice of $\tilde W$, the charge cancelation condition which
takes into account the discrete charge is as follows: if we regard
the Wilson line charge $X$ as an element of
$\Hom(\TT_\Lambda,U(1))$, then its restriction to $\ker V_\TT^t$
must have the form $\nu\circ \tilde W$ for some
$\nu\in\Hom(\TT_\M,U(1))$.

The choice of $\tilde W$ is constrained by the requirement that for any Wilson line ending on the boundary we must have
$$
\exp(2\pi i \sq(X))=1.
$$
This condition also implies that Wilson lines ending on the boundary
are mutually local and ensures that the subgroup $\sM\subset \D$ spanned by their charges is isotropic.

Boundary Wilson lines are restrictions of bulk Wilson lines. As before, charges of boundary Wilson lines take values in the quotient group $\D/\sM$. The tensor product of Wilson lines corresponds to the group operation in $\D/\sM$. The computation of the boundary associator and the semi-braiding proceeds in the same way as in section 5.2, the only difference being that now certain boundary Wilson lines are trivial because they can be screened by order operators $\exp(i\nu\varphi)$ on the boundary. The formulas of section 5.2 for the boundary associator and the semi-braiding remain valid.

%\section{Boundary-bulk map}\label{sec:bdrybulk}

\section{Surface operators}
\label{sec:surface}

\subsection{The folding trick}

A surface operator in a 3d TFT is a defect of codimension one. Such defects can be regarded as objects of a monoidal 2-category \cite{Kap:ICM}. Objects of this 2-category are surface operators themselves, morphisms are line operators sitting at the junction of two surface operators, and 2-morphisms are local operators sitting at the junction of two line operators. The monoidal structure arises from fusing surface operators placed next to each other.

Given a surface operators in a 3d TFT ${\mathcal A}$, we may construct a boundary condition in a 3d TFT $\cA\times \cA^*$, where $\cA^*$ denotes the parity reversal of $\cA$. This is accomplished by ``folding'' the worldvolume at the location of the surface operator. Using this folding trick, the 2-category of surface operators in the theory $\cA$ may be identified with the 2-category of boundary conditions in the theory $\cA\times \cA^*$.

If $\cA$ is an abelian Chern-Simons theory corresponding to the bilinear form $K$, then $\cA\times\cA^*$ is the abelian Chern-Simons theory corresponding to the bilinear form $K\oplus (-K)$. Thus the results obtained so far allow us to classify surface operators in an arbitrary abelian Chern-Simons theory and determine the 2-category structure on them. The monoidal structure is a new ingredient which requires a separate study.

Note that the form $K\oplus (-K)$ on the lattice $\Lambda\oplus\Lambda$ has signature zero, so there is no obstruction for the existence of topological boundary conditions in such a theory. In fact, in any such theory there are two obvious choices of a Lagrangian sublattice in $\Lambda\oplus\Lambda$: the diagonal one
$$
\Lambda_+=\left\{(\lambda,\lambda)\vert \lambda\in\Lambda\right\}
$$
and the anti-diagonal one
$$
\Lambda_-=\left\{(\lambda,-\lambda)\vert \lambda\in\Lambda\right\}.
$$
The diagonal boundary condition corresponds to the ``trivial'' or ``invisible'' surface operator whose insertion is equivalent to no surface operator at all. Such a surface operator is the identity object in the monoidal 2-category of surface operators.  The anti-diagonal surface operator is characterized by the fact that the gauge field changes its sign when one goes across the insertion surface.

The boundary condition for the $U(1)^{2m}$ theory considered in sections 4.4 and 5.3 corresponds to the trivial surface operator in the $U(1)^m$ theory. Note that while we have a whole family of such operators differing by a choice of the splitting $R$, they all appear to be isomorphic, since their physical properties (e.g. the associator for the line operators and the semi-braiding) do not seem to depend on $R$. To demonstrate this more formally, one needs to construct an invertible morphism between surface operators corresponding to different choices of $R$. From the physical viewpoint such a morphism and its inverse are boundary-changing line operators between the corresponding boundary conditions in the folded theory $\cA\times \cA^*$. Since we have not studied the composition of line operators sitting at the junction of different boundary conditions, we leave this for future work.

\subsection{Invertible surface operators and quantum equivalences}

In this section we will apply what we have learned so far to the problem of classification of abelian Chern-Simons theories on the quantum level. It was noticed in \cite{Witten:sl2action} that certain abelian Chern-Simons theories with a nontrivial $K$ are nevertheless trivial on the quantum level (isomorphic to the trivial theory). The classification problem for abelian Chern-Simons theories was studied in detail in \cite{BM}, both for even and odd lattices. According to \cite{BM}, two Chern-Simons theories are considered equivalent if their spaces of states on any Riemann surface are isomorphic as projective representations of the mapping class group. For even lattices the main result of \cite{BM} is that equivalence classes of abelian Chern-Simons theories are classified by the following data:
\begin{itemize}
\item Signature $\sigma$ of the bilinear form $K$ modulo 24;
\item The discriminant group $\D=\Lambda^*/K(\Lambda)$;
\item The quadratic function $\sq$: $\D \ra\QQ/\ZZ$ derived from $K$.
\end{itemize}
The analog of this for odd lattices is a bit more complicated; we will not discuss this case since it corresponds to spin Chern-Simons theories rather than regular 3d TFTs.

An example of a pair of even Chern-Simons theories which are equivalent on the quantum level but not classically
is given by positive-definite even unimodular lattices
$\Gamma_{16}$ and $\Gamma_8\oplus
\Gamma_8,$ where $\Gamma_{16}$ is the weight lattice of $Spin(32)/\ZZ_2$, and $\Gamma_8$ is the root lattice of the Lie algebra $E_8$.  These lattices are not linearly equivalent (i.e. there is no integral linear transformation $M$ with $\det M=\pm 1$ which establishes an isomorphism of the corresponding symmetric bilinear forms), but both have $\sigma=16$ and trivial $\D$ and $\sq.$

The classification scheme proposed in \cite{BM} identifies 3d TFTs which have isomorphic spaces of states on an arbitrary Riemann surface. It leaves open a possibility that other observables (e.g. line operators or surface operators) may distinguish theories equivalent in this sense. A better definition of equivalence which ensures that all topological observables coincide is based on the notion of a {\rm duality wall} \cite{GW,KapTi}. If two theories are truly equivalent, there should be a codimension-1 topological defect between them which implements the duality transformation on the fields. This defect enables one to map any observable in one theory to an observable in the other theory, so that correlators are preserved. The map is defined by inserting the codimension-1 defect on the boundary of a tubular neighborhood of the observable that one wishes to dualize. This is the reason for the name ``duality wall''. In three dimensions a duality wall is a surface operator which has an inverse. From the mathematical viewpoint, 3d TFTs form a 3-category, and a duality wall is an invertible morphism between two objects of this 3-category.

We will now show that a duality wall exists between any two abelian Chern-Simons theories with even $K$ for which the following data coincide:
\begin{itemize}
\item Signature $\sigma$ of the bilinear form $K$;
\item The discriminant group $\D$;
\item The quadratic function $\sq:\D\ra\QQ/\ZZ$ derived from $K$.
\end{itemize}
Note that this is almost the same data as in \cite{BM}, except that signatures are required to be identical rather than equal modulo 24. This is the best result one could possibly hope for. Indeed, since the properties of bulk line operators are described by $\D$ and $\sq$, Chern-Simons theories for which these data are different cannot be equivalent. Further, if the signatures are not identical, then no topological surface operator (invertible or not) can be constructed between such theories. Indeed, after performing the folding trick, the problem of finding a surface operator between two theories with bilinear forms $K_1,K_2$ becomes equivalent to the problem of finding a topological boundary condition for a theory with a bilinear form $K_1\oplus (-K_2)$. If $K_1$ and $K_2$ have different signatures, then $K_1\oplus (-K_2)$ has nonzero signature, and therefore does not admit any topological boundary conditions, as argued above. In other words, in the 3-category of 3d TFTs there are no morphisms between abelian Chern-Simons theories with different signatures.

To illustrate the difference between the classification scheme of \cite{BM} and that based on duality walls consider the theory based on a positive-definite even unimodular lattice $\Gamma_8^{\oplus 3}.$ This lattice has signature $24$, and therefore the theory is equivalent to the trivial one from the point of view of \cite{BM}. On the other hand, this theory, unlike the trivial theory, does not admit any topological boundary conditions and therefore is not equivalent to the trivial one from our viewpoint.

Let us proceed to prove the classification theorem. We will need a mathematical result due to Nikulin (Theorem 1.3.1 in \cite{Nik}) which says that two even lattices
$\Lambda_1$ and $\Lambda_2$ have isomorphic discriminant-bilinear
forms $\sb_1,\sb_2$ if and only if they are stably equivalent, i.e.
\be
\label{congr}\Lambda_1 \oplus {\cal S}_1 \cong \Lambda_2 \oplus
{\cal S}_2\ee where ${\cal S}_1$ and ${\cal S}_2$ are even
unimodular lattices. The symbol $\cong$ in (\ref{congr}) stands
for linear equivalence.

If $\Lambda_1$ and $\Lambda_2$ have the same signature, ${\cal S}_1$ and ${\cal S}_2$ also have the same signature. Since we can always replace ${\cal S}_1$ with ${\cal S}_1'={\cal S}_1\oplus (-{\cal S}_2)$ and ${\cal S}_2$ with ${\cal S}_2'={\cal S}_2\oplus (-{\cal S}_2)$, we may assume without loss of generality that ${\cal S}_1$ and ${\cal S}_2$ have signature zero. Another well-known mathematical fact is that any even unimodular lattice with zero signature is isomorphic to a sum of several copies of the rank-two lattice $U$, whose symmetric bilinear form is
$K_U=\left(\begin{tabular}{cc}$0$ & $1$\\ $1$ &
$0$\end{tabular}\right)$. Putting all this together, we see that  if $\Lambda_1$ and $\Lambda_2$ have the same $\sigma$, $\D$, and $\sq$, then there exists an integral matrix
$M$ with $det(M)=\pm 1$ such that
\be \label{stable}
K_1\oplus K_U^{\oplus m_1}=M^T(K_2\oplus K_U^{\oplus m_2})M
\ee
This implies that the theories defined by the bilinear forms $K_1'=K_1\oplus K_U^{\oplus m_1}$ and $K_2'=K_2\oplus K_U^{\oplus m_2}$ are equivalent and related by the change of variables in the path-integral which sets $A_2'=M A_1'$. The duality wall between them is the surface operator defined by the condition  that the restriction of $A_2'$ is equal to the restriction of $MA_1'$.

To complete the proof it is sufficient to show that for any even symmetric bilinear form $K$ and any integer $m$ the Chern-Simons theories corresponding to $K$ and $K\oplus K_U^{\oplus m}$ are equivalent. In fact, it is sufficient to show this for the case $K=0$ and $m=1$: given a duality wall between $K_U$ and the trivial theory one may construct a duality wall between $K\oplus K_U^{\oplus m}$ and $K$ by combining $m$ copies of the former duality wall with the ``invisible'' wall between $K$ and  $K$.

The fact that the Chern-Simons theory corresponding to $K_U$ is trivial was explained in \cite{Witten:sl2action}. We would like to demonstrate this more formally by exhibiting a duality wall between the theory $K_U$ and the trivial theory. Such a wall is a boundary condition for the theory $K_U$. An obvious choice of a Lagrangian subgroup in $U(1)\times U(1)$ is given by
$$
P=\begin{pmatrix} 1 \\ 0 \end{pmatrix}.
$$
This means that the restriction of the gauge field to the boundary has the form $(A_1,0)$, where $A_1$ is an arbitrary $U(1)$ gauge field. The boundary gauge group is $U(1)$. We can take the splitting $R$ to be $R=P$.  Let us denote by ${\cal O}_{\emptyset;U}$ the resulting surface operator between the theory $K_U$ and the trivial one. We need to show that it is invertible. The inverse is the same surface operator, but parity-reversed; we will denote it ${\cal O}_{U;\emptyset}$. We need to show that the surface operators ${\cal O}_{\emptyset;U}$ and ${\cal
O}_{U;\emptyset}$ satisfy
\be \label{trivial} {\cal
O}_{\emptyset;U}{\cal O}_{U;\emptyset}=\Id_{\emptyset},\quad {\cal
O}_{U;\emptyset}{\cal O}_{\emptyset;U}=\Id_U,\ee
where $\Id_{\emptyset}$ i the invisible surface operator in the trivial theory, and $\Id_U$ is the invisible surface operator in the theory $K_U$.

To prove the first equality in (\ref{trivial}) we need to study the Chern-Simons theory defined by $K_U$ on an interval $[0,L]$ times $\Sigma$, where $\Sigma$ is an arbitrary Riemann surface. The boundary conditions are given by $P$ on both ends of the interval. We use the fact that Chern-Simons theory on $M_3=\Sigma\times [0,L]$ with such boundary conditions reduces to the $U(1)$ $BF$-type theory with the action
\begin{equation}\label{BFtrivial}
S=\frac{i}{2\pi}\int_\Sigma \varphi \ dA,
\end{equation}
where $\varphi$ is a periodic scalar with period $2\pi$. This reduction was explained in section 5.4. Such a theory is trivial for any $\Sigma$ (its partition function is one), which proves the first equality in (\ref{trivial}).

To prove the second equality, we need to consider  the theory $K_U$ on an arbitrary 3-manifold $M_3$ and with an insertion of ${\cal
O}_{U;\emptyset}{\cal O}_{\emptyset;U}$ along a Riemann surface $\Sigma$ embedded into $M$. This means that we excise a region $H_3=\Sigma\times [0,L]$ and insert
${\cal O}_{U;\emptyset}$ at $\Sigma\times\{0\}$ and ${\cal O}_{\emptyset;U}$ at $\Sigma\times \{L\}$.  The theory $K_U$ assigns to $\Sigma$ a one-dimensional Hilbert space $\cL$ \cite{Witten:sl2action}.  The boundary of the 3-manifold $M_3\backslash H_3$ is a disjoint sum of two copies of $\Sigma$ with opposite orientation, so the theory $K_U$ attaches to it a vector in the space $\cL\otimes\cL^*=\End(\cL)$. Let us denote this vector $\Phi(M_3;H_3)$. The partition function of the theory $K_U$ on $M_3$ is given by the inner product
$$
\langle \Id_\Sigma\vert \Phi(M_3,H_3)\rangle,
$$
where $\Id_\Sigma$ is the identity element in $\End(\cL)$ (it represents the invisible surface operator inserted at $\Sigma$).
On the other hand, the partition function of the theory $K_U$ on $M_3$ with an insertion of ${\cal O}_{U;\emptyset}{\cal O}_{\emptyset;U}$ along $\Sigma$ is given by the inner product
$$
\langle \Psi_\Sigma\otimes\Psi_\Sigma^*\vert \Phi(M_3,H_3)\rangle,
$$
where $\Psi_\Sigma\in\cL$ is the boundary state corresponding to the boundary condition ${\cal O}_{U;\emptyset}$. Since the Hilbert space $\cL$ is one-dimensional, the state
$\Psi_\Sigma\otimes\Psi_\Sigma^*$ is equal to $\Id_\Sigma$ times the norm of $\Psi_\Sigma$. The latter norm is equal to the partition function of the BF-theory (\ref{BFtrivial}) on $\Sigma$, i.e. $1$. Hence the partition function of the theory $K_U$ on $M_3$ coincides with the partition function of the same theory with an insertion of ${\cal
O}_{U;\emptyset}{\cal O}_{\emptyset;U}$. This proves the second equality in (\ref{trivial}).

 \section{Concluding remarks}\label{sec:concl}

 In this paper we studied topological boundary conditions in an arbitrary abelian Chern-Simons theory and argued that they are classified by subgroups of the discriminant group $\D$ which are Lagrangian with respect to the quadratic function $\sq:\D\ra\QQ/\ZZ$. Given such a Lagrangian subgroup $\sM$, boundary line operators are labeled by elements of the finite group $\D/\sM$. The monoidal category of boundary line operators is a twisted version of the category of $\D/\sM$-graded vector spaces, where the twist is given by a certain canonical element in $H^3(\D/\sM,\QQ/\ZZ)$ which we explicitly described. Furthermore, we showed that for any Lagrangian $\sM$  there is a monoidal functor (which we called the semi-braiding) from the category of bulk line operators to the Drinfeld center of the category of boundary line operators.

 These results raise the question how to describe the full 2-category of boundary conditions for an abelian Chern-Simons theory. By analogy with the Rozansky-Witten model \cite{KRS,KR} one may propose the following conjecture. Let $\sM$ be a Lagrangian subgroup of $\D$, and let $\cC_\sM$ be the corresponding monoidal category of boundary line operators. To any other boundary condition one may attach a module category over $\cC_\sM$, namely, the category of line operators sitting at the junction of the chosen boundary condition and $\sM$. Axioms of 3d TFT imply this map extends to a functor from the 2-category of boundary conditions to the 2-category of module categories over
 $\cC_\sM$. It is natural to conjecture that this functor is an equivalence. In other words, different choices of $\sM$ give 2-Morita-equivalent monoidal categories $\cC_\sM$. The Drinfeld centers of all these monoidal categories are equivalent, and presumably the semi-braiding functors establishes an equivalence between these Drinfeld centers and the braided monoidal category of bulk line operators. Note that if we take $\sM$ be isotropic rather than Lagrangian, then the formula for the element in $H^3(\D/\sM,\QQ/\ZZ)$ still makes sense, so the monoidal category of $\D/\sM$-graded vectors spaces twisted by this cocycle is well-defined, but its Drinfeld center is ``too large'' and is not equivalent to the category of bulk line operators. Presumably the role of the coisotropic condition is to ensure that the Drinfeld center of the category $\cC_\sM$ is of the right size.

In this paper we only briefly discussed surface operators in abelian Chern-Simons theory. It would be interesting to understand how to fuse surface operators, i.e. to understand the monoidal structure on the 2-category of surface operators. It would be even more interesting to extend the analysis to nonabelian Chern-Simons theories. Nonabelian Chern-Simons theories with a simple gauge group do not seem to admit topological boundary conditions, but they have interesting surface operators whose properties are not understood.

Recently higher categorical structures in abelian Chern-Simons theory have been studied from a rather different standpoint by Freed, Hopkins, Lurie and Teleman \cite{FHLT}. These authors consider the situation when the signature of $K$ is not necessarily zero, and the theory itself is regarded as an ``anomalous'' 3d TFT which sits on a boundary of a 4d TFT. It would be very interesting to understand the relationship between our work and \cite{FHLT} in the case when $K$ has zero signature.

\section*{Acknowledgments} We would like to thank D. Freed, A. Kitaev, J. Lurie, G. Moore, V. Ostrik, and L. Rozansky for useful discussions and advice. We are grateful to the Aspen Center for Physics for an excellent working atmosphere. This work was supported in part by the DOE grant DE-FG02-92ER40701.

\section*{Appendix A. Braided monoidal categories}
A monoidal category ${\cal V}$ is a category with a covariant bi-functor (tensor product)
$\otimes: \, {\cal V}\times {\cal V} \mapsto {\cal V}$ and a distinguished object $1.$ In addition, for any three objects $U,V,W$ one is given an isomorphism
$$a_{U,V,W}: \, (U\otimes V)\otimes W \mapsto U\otimes (V \otimes W),$$
and for any object $U$ one is given a pair of isomorphisms
$$r_U: U\otimes 1 \mapsto U, \quad l_U: 1\otimes U \mapsto U $$
such that the following pentagon and triangle diagrams commute:

\vspace{2cm}

\xymatrixcolsep{0.05pc}
\centerline{\xymatrix{
& (U\otimes V) \otimes (W\otimes X)\ar[rd]^{a_{U,V,W\otimes X}} & \\
\bigl((U\otimes V)\otimes W\bigr)\otimes X \ar[ru]^{a_{U\otimes V,W,X}}  \ar[d]^{a_{U,V,W}\, \otimes id_X} & & U\otimes\bigl(V\otimes (W\otimes X)\bigr)\\
\bigl(U\otimes (V\otimes W)\bigr)\otimes X \qquad \qquad \ar[r]^-{a_{U,V\otimes W,X}}  & &
U\otimes\bigl((V\otimes W)\otimes X\bigr)\ar[u]_{id_U\otimes \, a_{V,W,X}}}}

\vspace{2cm}

\xymatrixcolsep{2pc}
\centerline{\xymatrix{
(U\otimes 1)\otimes V \ar[rd]_{r_U\, \otimes \, id_V} \qquad \ar[r]^-{a_{U,1,V}} & & U\otimes (1\otimes V) \ar[ld]^{id_U \otimes \, l_V} \\
&U\otimes V &}}

\vspace{2cm}

The morphism $a_{U,V,W}$ is called the associator. If both $a_{U,V,W}$ and $r_U,l_U$ are identities, the monoidal category is called strict. In the cases of interest to us $r_U$ and $l_U$ are identities (i.e. we are dealing with monoidal categrories with a strict identity), but the associator may be nontrivial.

Let $\G$ be a finite abelian group, and let us consider the category of finite-dimensional $\G$-graded complex vector spaces. A simple objects in this category is a one-dimensional vector space $\CC_X$ labeled by a particular element $X\in\G$; its endomorphism algebra is $\CC$ for any $X\in\G$. Let us define the tensor product of simple objects $\CC_X$ and $\CC_Y$ to be $\CC_{X+Y}$. The associator is an element of $\CC^*$
$$a_{X,Y,Z}=\exp\bigl(2\pi i h(X,Y,Z)\bigr)\in \CC^*,\quad X,Y,Z \in \G.$$
Regarding $h(X,Y,Z)$ as an $\CC/\ZZ$-valued function on $\G\times\G\times\G$, we may write the pentagon identity as follows:
$$h(X,Y,Z+W)+h(X+Y,Z,W)=h(Y,Z,W)+h(X,Y+Z,W)+h(X,Y,Z).$$
The triangle identity reads
$$h(X,0,Z)=0.$$
We will denote such a monoidal category $(\Vect_\G,h)$. One can show that $(\Vect_\G,h)$ is equivalent to $(\Vect_\G,h')$ if and only if there exists a function
$k:\G\times\G\ra\CC/\ZZ$ such that
$$
(h'-h)(X,Y,Z)=k(Y,Z)-k(X+Y,Z)+k(X,Y+Z)-k(X,Y).
$$
In such a case we will say that $h$ and $h'$ differ by a coboundary.

A braided monoidal category is equipped with an additional structure, a family of braiding isomorphisms
$$c_{U,V}: U\otimes V \mapsto V \otimes U.$$
Braiding is required to be compatible with the tensor product, by which we mean that the following two hexagon diagrams must commute:

\xymatrixcolsep{2pc}
\centerline{\xymatrix{
& A\otimes (B\otimes C)\ar[ld]_{a^{-1}_{A,B,C}} \ar@{=>}[r]^{c_{A,B\otimes C}} &(B\otimes C)\otimes A & \\
(A\otimes B)\otimes C \ar[rd]^{c_{A,B}\otimes id_C} & & &B\otimes(C\otimes A) \ar[lu]_{a^{-1}_{B,C,A}}\\
 & (B\otimes A)\otimes C \ar[r]^{a_{B,A,C}} &B\otimes(A\otimes C)\ar[ru]^{id_B \otimes c_{A,C}} &}}

\vspace{2cm}

\centerline{ \xymatrix{
& (U\otimes V)\otimes W\ar[ld]_{a_{U,V,W}} \ar@{=>}[r]^{c_{U\otimes V,W}} &W\otimes(U\otimes V) & \\
U\otimes (V\otimes W)\ar[rd]^{id_U\otimes c_{V,W}} & & &(W\otimes U)\otimes V \ar[lu]_{a_{W,U,V}}\\
 & U\otimes (W\otimes V) \ar[r]^{a^{-1}_{U,W,V}} &(U\otimes W)\otimes V\ar[ru]^{c_{U,W}\otimes id_V} &}}

\vspace{2cm}

Consider again the monoidal category $(\Vect_G,h)$. A braiding on this category is a $\CC^*$-valued function on $\G\times\G$,
$$c_{X,Y}=\exp\bigl(2\pi i s(X,Y)\bigr),\quad X,Y \in \G$$
and the hexagon identities become
\begin{align}\label{hexbulk1}
s(X,Y+ Z)-s(X,Y)-s(X,Z)&=h(Y,X,Z)-h(Y,Z,X)-h(X,Y,Z),\\ \label{hexbulk2}
s(X+Y,Z)-s(Y,Z)-s(X,Z)&=h(Z,X,Y)-h(X,Z,Y)+h(X,Y,Z).
\end{align}
Here we regard $s(X,Y)$ as an $\CC/\ZZ$-valued function.

It was shown in \cite{JS} that pairs $(h, s)$ and $(h^{\prime},
s^{\prime})$ correspond to equivalent braided monoidal categories if
their difference is a ``coboundary'', i.e.
\begin{align}
(h'-h)(X,Y,Z)&=k(Y,Z)-k(X+Y,Z)+k(X,Y+Z)-k(X,Y),\\
(s'-s)(X,Y)&=-k(X,Y)+k(Y,X),
\end{align}
where $k(X,Y)$ is an arbitrary $\CC/\ZZ$-valued function on $\G\times\G$.

\section*{Appendix B. Drinfeld center, semi-braiding and the boundary hexagon relations}
 The Drinfeld center of a monoidal category $\cC$ is a
(braided) monoidal category $\Z(\cC)$ whose objects are pairs
$(c,\chi_c)$, where $c$ is an object of $\cC$ and $\chi_c$ is a
family of natural isomorphisms $\chi_c(a): a\otimes c\ra c\otimes a$ for all
objects $a$ of $\cC$. Moreover, $\chi_c$ must be compatible with the monoidal structure on $\cC$ in the sense that the following diagram commutes:

\vspace{1cm}

\xymatrixcolsep{2pc} \centerline{ \xymatrix{ & (a\otimes b)\otimes
c\ar[ld]_{\mathbf{a}_{a,b,c}}
\ar@{=>}[r]^{\chi_{c}(a\otimes b)}
&c\otimes(a\otimes b) & \\
a\otimes (b\otimes c)\ar[rd]^{id_a\otimes \chi_c(b)} &
& &(c\otimes a)\otimes b
\ar[lu]_{\mathbf{a}_{c,a,b}}\\
 & a\otimes (c\otimes b) \ar[r]^{\mathbf{a}^{-1}_{a,c,b}} &
 (a\otimes c)\otimes b\ar[ru]^{\chi_c(a)
 \otimes id_b} &}}
 Here $\mathbf{a}_{a,b,c}$ denotes the associator in the category $\cC$.

Let us specialize to the case $\cC=(\Vect_\G,h)$. In this case we can regard $\chi_c$ as a function $\chi_c:\G\ra\CC/\ZZ$ and the commutativity of the above diagram is equivalent to the equation
\be \label{chi_cond}\chi_c(a+b)=\chi_c(a)+\chi_c(b)+h(c,a,b)-h(a,c,b)+h(a,b,c)\ee

We have mentioned above that the Drinfeld center of a monoidal category has a natural braided monoidal structure. Let us explain how the tensor product on objects of $\Z(\cC)$ looks like in the case $\cC=(\Vect_\G,h)$. The tensor product of two simple objects $(b,\chi_b)$ and $(c,\chi_c)$ in the Drinfeld center, $b,c\in\G$, is
defined as $(b+c,\chi_{b,c})$ with
\begin{equation}\label{drinfeld_tensor}
\chi_{b,c}(a)= \chi_b(a)+\chi_c(a)+h(b,a,c)-h(b,c,a)-h(a,b,c).
\end{equation}

In this paper we considered the following situation: we have a pair of monoidal categories $(\Vect_\D,h)$ and $(\Vect_{\D_0},h_0)$ such that $\D_0$ is a quotient of $\D$, and a function $s_{sb}:\D_0\times\D\ra\CC/\ZZ$. We would like to interpret this function as defining a strong monoidal functor from the former category to the Drinfeld center of the latter. This functor is supposed to map an object $B\in\D$ to an object $(b,s_{sb}(\cdot;B))$ of the Drinfeld center, where $b=\pi(B)$ and $\pi:\D\ra \D_0$ is the quotient map. The condition that this satisfies (\ref{chi_cond}) reads
$$
s_{sb}(a\plusd b;C)-s_{sb}(a;C)-s_{sb}(b;C)=h_0(c,a,b)-h_0(a,c,b)+h_0(a,b,c),
$$
where we denoted the group operation by $\plusd$ to agree with the notation in the rest of the paper. We will call this equation the first boundary hexagon relation.

Note that this hexagon relation is satisfied if we set $\D_0=\D$ and $s_{sb}(b;C)=s(b,c)$. This means that the category of bulk line operators naturally embeds into its own Drinfeld center. More generally, any braided monoidal category naturally embeds into its own Drinfeld center. If we regard a monoidal and a braided monoidal category as a categorification of an associative algebra and a commutative associative algebra respectively, this is analogous to the statement that a commutative algebra coincides with its own center.

A strong monoidal functor must preserve the tensor product of objects. Recalling the definition of the tensor product (\ref{drinfeld_tensor}) we conclude that the following equation must hold:
\be
\label{second}s_{sb}(a; B\plusd C)=s_{sb}( a; B)+s_{sb}( a; C)+
h_0(b, a,  c)-h_0( b,  c, a)-h_0( a, b, c). \ee
We will this equation the second boundary hexagon relation because it is equivalent to the commutativity of the following diagram:

\vspace{1cm}
\xymatrixcolsep{3pc} \centerline{\xymatrix{ & a\otimes (b\otimes
c)\ar[ld]_{\mathbf{a}^{-1}_0(a,b,c)}
\ar@{=>}[r]^{\mathbf{c}_{sb}(a;B\otimes C)}
&(b\otimes c)\otimes a & \\
(a\otimes b)\otimes c \ar[rd]^{~~\mathbf{c}_{sb}(a;B)\otimes id_c}
& & &b\otimes(c\otimes a)
\ar[lu]_{\mathbf{a}^{-1}_0(b,c,a)}\\
 & (b\otimes a)\otimes c \ar[r]^{\mathbf{a}_0(b,a,c)} &b\otimes(a\otimes c)\ar[ru]^{id_b
  \otimes \mathbf{c}_{sb}(a;C)} &}}

\vspace{2cm}
Here $\mathbf{c}_{sb}(a;B)=\exp(2\pi i s_{sb}(a;B))$ and $\mathbf{a}_0(a,b,c)=\exp(2\pi i h_0(a,b,c))$.

Again, the equation (\ref{second}) automatically holds if we set $\D_0=\D$ and $s_{sb}(b;C)=s(b,c)$, meaning that the natural embedding of the braided monoidal category $(\Vect_\D,h)$ into its own Drinfeld center preserves tensor products.

A strong monoidal functor $F$ between two monoidal categories ${\mathsf B}$ and $\mathsf C$ is more than merely a functor which preserves tensor products. For any two objects $A,B$ of ${\mathsf B}$ one must be given an isomorphism $\Phi_{A,B}: F(A\otimes B)\ra F(A)\otimes F(B)$ satisfying a compatibility relation with the associators. One should also be given an isomorphism $F(\Id)\ra \Id'$ between identity objects satisfying a compatibility condition. In our case the latter isomorphism is the identity, and the corresponding compatibility condition is trivially satisfied. On the other hand, $\Phi_{A,B}$ is a nontrivial phase $\exp(2\pi i \phi(A,B))$ satisfying
 \be \label{mono}
 h_0( a, b,c)-h(A, B, C)= \phi( A, B+  C)-\phi( A, B)-\phi(A+  B, C)+\phi(B, C).\ee
In our case this phase turns out to be
$$\phi(A,B)=\half \sum_{\c}a_\c m_\c \left[{B_\c \over m_\c}\right]
(\tv_\c K^{-1} \tv_\c)+\sum_{\c>\b}a_\b m_\c \left[{B_\c \over
m_\c}\right] (\tv_\b K^{-1} \tv_\c).$$

Finally note that the pentagon equation and the two boundary hexagon relations continue to hold if $(s_{sb},h_0)$ is replaced with $(s'_{sb},h'_0)$
such that
\begin{align}
s'_{sb}(a;B)-s_{sb}(a;B)&=-k(a,b)+k(b,a),\\
h'_0(a,b,c)-h_0(a,b,c)&=k(a,b+c)-k(a+b,c)+k(b,c)-k(a,b),
\end{align}
where $k:\D_0\times\D_0\ra\CC/\ZZ$ is an arbitrary function. Such a transformation corresponds to replacing $(\Vect_{\D_0},h_0)$ with an equivalent monoidal category, and modifying appropriately the semi-braiding functor (composing it with the equivalence).


\begin{thebibliography}{99}

\bibitem{Kap:ICM} A.~Kapustin, ``Topological Field Theory, Higher Categories, and Their Applications,''
  arXiv:1004.2307 [math.QA].

\bibitem{BM}{D. Belov, G. Moore,``Classification of abelian spin Chern-Simons theories'', hep-th/0505235.
}

\bibitem{GW} D.~Gaiotto and E.~Witten, ``S-Duality of Boundary Conditions In N=4 Super Yang-Mills Theory,''
  arXiv:0807.3720 [hep-th].

\bibitem{KapTi} A.~Kapustin and M.~Tikhonov, ``Abelian duality, walls and boundary conditions in diverse dimensions,''
  JHEP {\bf 0911}, 006 (2009)
  [arXiv:0904.0840 [hep-th]].

\bibitem{JS}{A.~Joyal, R.~Street, ``Braided tensor categories'',
Adv. ~Math. ~102, 20-78, 1983.}

\bibitem{BK} B.~Bakalov and A.~A.~Kirillov, ``Lectures on tensor categories and modular functors,'' American Mathematical Society, 2000.

\bibitem{Sti} S.~D.~Stirling, ``Abelian Chern-Simons theory with toral gauge group, modular tensor categories, and group categories,''
  arXiv:0807.2857 [hep-th].

\bibitem{KRS} A.~Kapustin, L.~Rozansky and N.~Saulina, ``Three-dimensional topological field theory and symplectic algebraic geometry I,''
  Nucl.\ Phys.\  B {\bf 816}, 295 (2009)
  [arXiv:0810.5415 [hep-th]].


 \bibitem{Witten}{E. Witten, ``Quantum field theory and the Jones polynomial'', Com.~Math.~Phys.~121, 351-399.}

\bibitem{Witten:sl2action} E.~Witten, ``SL(2,Z) action on three-dimensional conformal field theories with Abelian symmetry,'' arXiv:hep-th/0307041.

\bibitem{Nik}{V.~V.~Nikulin, ``Integral symmetric bilinear forms and their applications'',
Math. USSR. Izvestija Vol. 14,  p. 103.}

\bibitem{KR} A.~Kapustin and L.~Rozansky,``Three-dimensional topological field theory and symplectic algebraic geometry II,''
  arXiv:0909.3643 [math.AG].

\bibitem{FHLT} D.~S.~Freed, M.~J.~Hopkins, J.~Lurie and C.~Teleman, ``Topological Quantum Field Theories from Compact Lie Groups,''
  arXiv:0905.0731 [math.AT].


\end{thebibliography}
\end{document}